\pdfoutput=1
\documentclass{article} %
\usepackage{amsmath}
\usepackage{amssymb}
\usepackage{booktabs}
\usepackage{color}
\usepackage[colorlinks=true]{hyperref}
\usepackage{enumerate} %
\usepackage{epsfig} %
\usepackage{epstopdf} %
\usepackage{gensymb} %
\usepackage{graphbox} %
\usepackage{graphicx}
\usepackage{layouts}
\usepackage[misc]{ifsym}
\usepackage{nicefrac} %
\usepackage{paralist}
\usepackage{subcaption}
\usepackage{tikz}
\usepackage{rotating}
\usepackage{todonotes}
\usepackage{hyperref}
\usepackage{fontawesome}

\hypersetup{urlcolor=blue, citecolor=red}
\allowdisplaybreaks

\usetikzlibrary{arrows.meta,calc,decorations.markings}

\newcommand{\figurewidth}{0.9\linewidth}

\DeclareMathOperator*{\smoothstep}{smoothstep}
\DeclareMathOperator*{\brightness}{brightness}

\newcommand{\C}{}
\newcommand{\sout}[1]{}

\newboolean{iscleanversion}
\setboolean{iscleanversion}{true}

\title
{Limited-Angle Tomography Reconstruction via Deep End-To-End Learning on Synthetic Data} %

\author{Thomas Germer\footnote{Heinrich Heine University Düsseldorf, corresponding author \href{mailto:thomas.germer+htc@hhu.de}{\faEnvelope}}
    \and
    Jan Robine\footnote{Technical University of Dortmund}
    \and
    Sebastian Konietzny\textsuperscript{\textdagger}
    \and
    Stefan Harmeling\textsuperscript{\textdagger}
    \and
    Tobias Uelwer\textsuperscript{\textdagger}
}

\makeatletter
\def\@maketitle{%
    \begin{center}%
        \let \footnote \thanks
        {\LARGE \@title \par}%
        \vskip 2em%
        {\large
            \begin{tabular}[t]{c}%
                \@author
            \end{tabular}\par}%
    \end{center}%
    \par
    \vskip 1em}
\makeatother

\begin{document}
\maketitle

\begin{abstract}
    Computed tomography (CT) has become an essential part of modern science and medicine. A CT scanner consists of an X-ray source that is spun around an object of interest. On the opposite end of the X-ray source, a detector captures X-rays that are not absorbed by the object. The reconstruction of an image is a linear inverse problem, which is usually solved by filtered back projection. However, when the number of measurements is small, the reconstruction problem is ill-posed. This is for example the case when the X-ray source is not spun completely around the object, but rather irradiates the object only from a limited angle. To tackle this problem, we present a deep neural network that is trained on a large amount of carefully-crafted synthetic data and can perform limited-angle tomography reconstruction even for only 30\degree or 40\degree sinograms. With our approach we won the first place in the Helsinki Tomography Challenge 2022.
\end{abstract}

\newcommand{\sinlen}{m}
\newcommand{\objlen}{n}

\section{Introduction}\label{sec:introduction}
\begin{figure}[!htb]
    \begin{center}
        \includegraphics[width=\figurewidth]{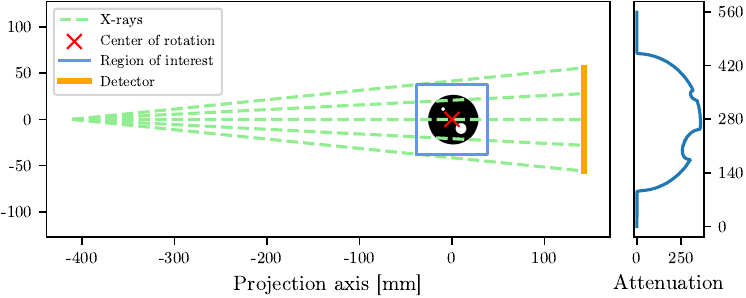}
        \caption{X-ray tomography experimental setup for the Helsinki Tomography Challenge 2022 \cite{meaney2022helsinki}.}
        \label{fig:setup}
    \end{center}
\end{figure}

Computed tomography (CT) \cite{hounsfield1973computerized} is an imaging technique which has become indispensable in modern science and medicine to visualize the interior of objects of interest.
To this end, a detector measures the intensity $I_1$ of X-rays after they have passed through an object, as shown in Figure~\ref{fig:setup}.

While traveling along a line $L$, the initial intensity $I_0$ is attenuated exponentially as it passes through matter with attenuation coefficient $\mu$ at point $p$, following the Beer-Lambert law:
\begin{equation}
  I_1 = I_0 e^{- \C{\int_L} \mu(p) \C{\mathrm{d} p}}.
\end{equation}
From a mathematical viewpoint, it is easier to work in a linear space. Therefore, we will instead consider the form
\begin{equation}
  \int_L \mu(p) \mathrm{d} p = -\log\left(\frac{I_1}{I_0}\right)
\end{equation}
throughout this article.

\begin{figure}[!htb]
    \centering
    \begin{subfigure}[t]{.3\textwidth}
        \centering
        \includegraphics[height=2.5cm]{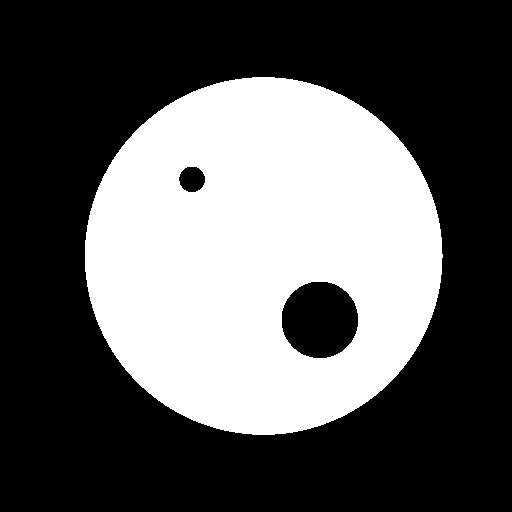}
        \captionsetup{justification=centering}
        \caption{A disk with two holes of different sizes.}
        \label{fig:objectofinterest}
    \end{subfigure}
    \hspace{5mm}
    \begin{subfigure}[t]{.3\textwidth}
        \centering
        \includegraphics[height=2.5cm]{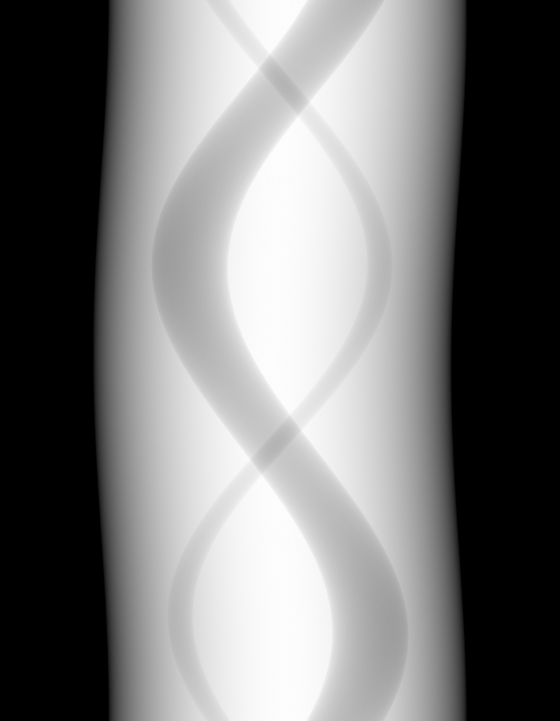}
        \captionsetup{justification=centering}
        \caption{Sinogram of the disk.}
        \label{fig:sinogram}
    \end{subfigure}
\caption{Example object and its corresponding sinogram.}
\end{figure}

For a given rotation angle of the object of interest, a set of line integrals is obtained (Figure~\ref{fig:setup}, on the right side).
Depending on the experimental setup, the object of interest is then rotated around its center, or alternatively, the measurement apparatus is rotated around the object, resulting in the rows of a sinogram (Figure~\ref{fig:sinogram}).

By discretizing the line integrals, the process of obtaining the sinogram $y \in \mathbb{R}^{\sinlen}$ from an object of interest $x \in \mathbb{R}^{\objlen}$ can be expressed as a linear system
\begin{equation}
  \label{eq:problem}
  y = A x + \epsilon,
\end{equation}
where $\epsilon$ is additive measurement noise.
Each row of the matrix describes the influence of a specific X-ray as a weighted sum over the attenuation coefficients of the object $x$. The weights depend on the path of the X-ray through space and are mostly zero since a single ray only travels through a small part of the object. Hence, the matrix $A$ is sparse.

\subsection{Problem Description}
The heart of computed tomography is the reconstruction algorithm that solves a linear inverse problem to obtain the image $x$ of the object from the sinogram $y$.
For a good reconstruction, it is preferable to capture measurements from a multitude of directions.
However, that is not always possible in practice due to constraints on the measurement setup or due to radiation limits, as it is the case in digital breast tomosynthesis~\cite{niklason1997digital}, or dental X-ray imaging~\cite{doi:10.1137/1.9781611972344}.
For this reason, we are considering the problem  of limited angle tomography, where the sinogram only covers a small range of angles.

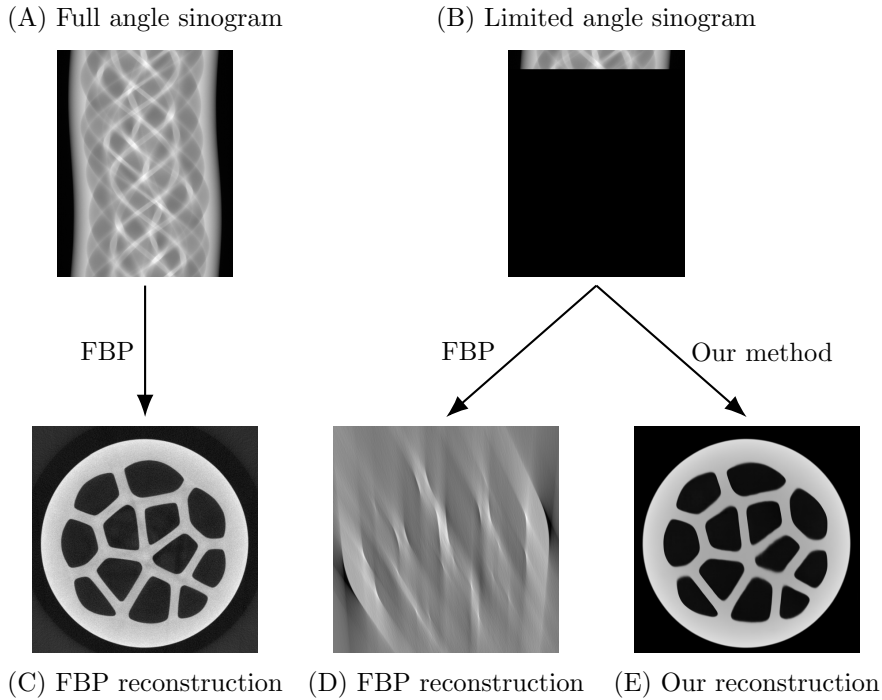
\begin{figure}
    \centering
    \begin{tikzpicture}
            \node[label=above:{(A) Full angle sinogram}](fullsinogram) at (-3, 0) {\includegraphics[height=3cm]{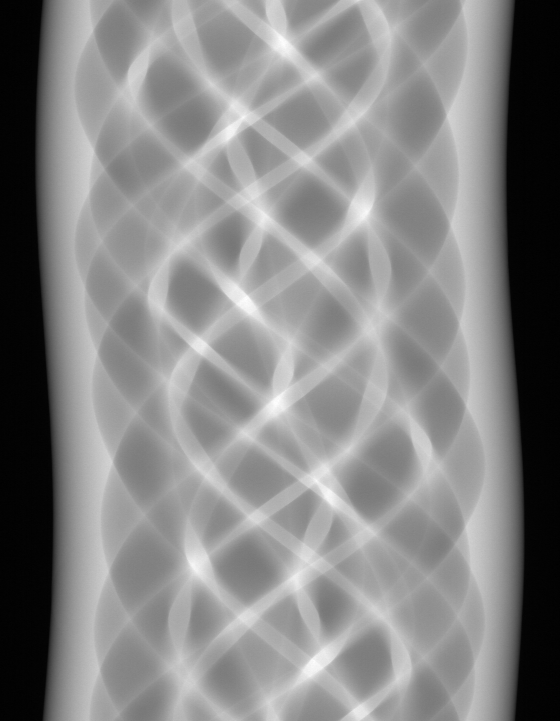}};
            \node[label=above:{(B) Limited angle sinogram}](limitedsinogram) at (3, 0) {\includegraphics[height=3cm]{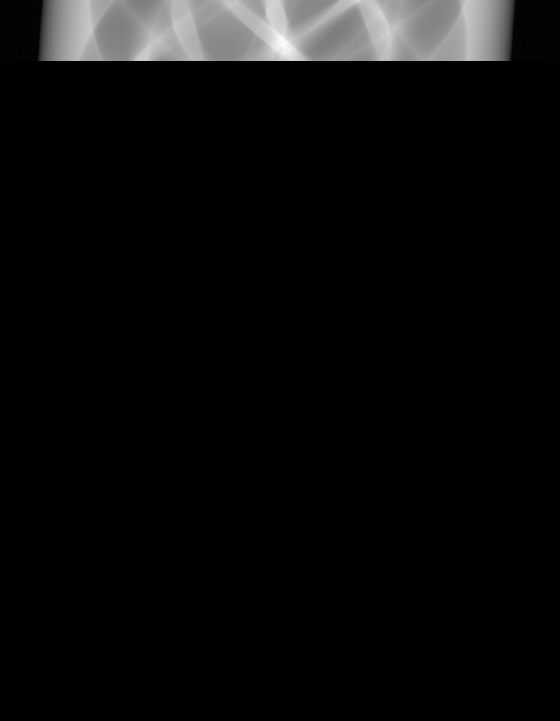}};
            \node[label=below:{(C) FBP reconstruction}](fullfbpreconstruction) at (-3, -5) {\includegraphics[height=3cm]{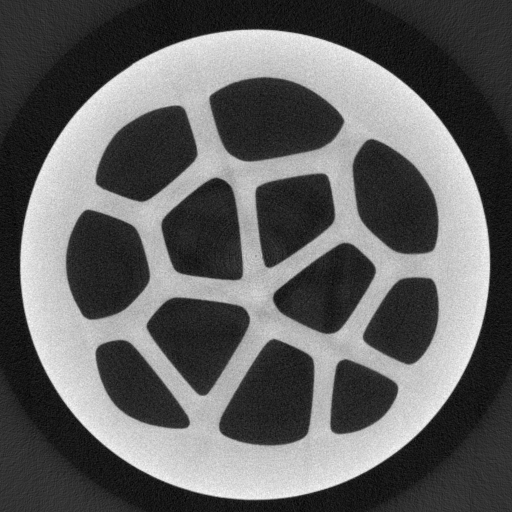}};

            \node[label=below:{(D) FBP reconstruction}](limitedfbpreconstruction) at (1, -5) {\includegraphics[height=3cm]{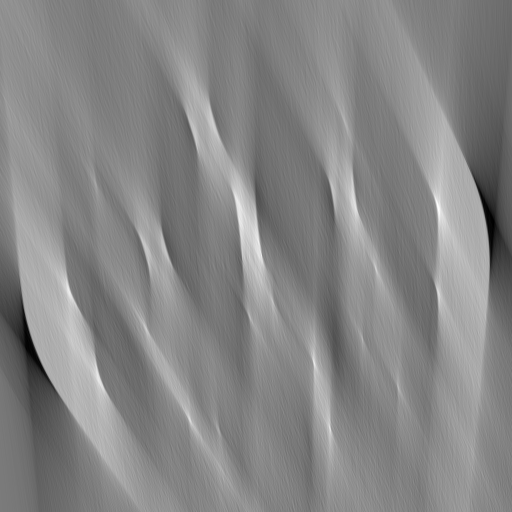}};
            \node[label=below:{(E) Our reconstruction}](ourreconstruction) at (5, -5) {\includegraphics[height=3cm]{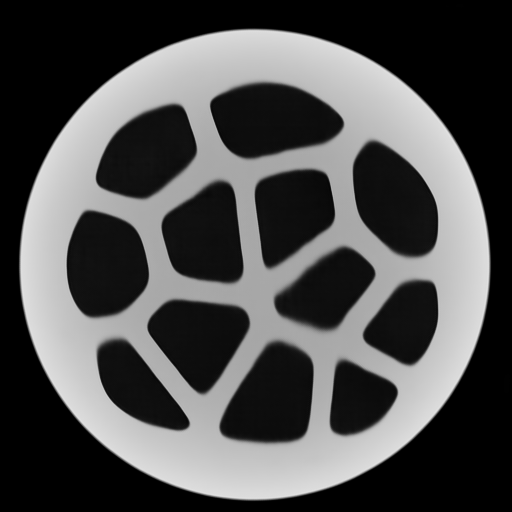}};

             \draw [thick, -{Latex[length=3mm]}] ([yshift=0cm]fullsinogram.south) -- (fullfbpreconstruction.north);
              \draw [thick, -{Latex[length=3mm]}] ([yshift=0cm]limitedsinogram.south) -- (limitedfbpreconstruction.north);
               \draw [thick, -{Latex[length=3mm]}] ([yshift=0cm]limitedsinogram.south) -- (ourreconstruction.north);
               \node at (-3.5,-2.5) {{FBP}};
               \node at (1.3,-2.5) {{FBP}};
                \node at (5.2,-2.5) {{Our method}};
    \end{tikzpicture}
\caption{\C{Given a full sinogram (A), the filtered back projection (FBP) algorithm can reconstruct the original image (C). However, when the angular range of the sinogram is limited (B), the FBP reconstruction (D) shows severe artifacts. Our method (E) can reconstruct the image from the limited angle sinogram.}}
\label{fig:limited_angle_reconstruction}
\end{figure}

\subsection{Related Work}

For full-angle CT, reconstruction can be performed by traditional algorithms like the filtered back projection (FBP) algorithm~\cite{feldkamp1984practical}. However, these algorithms fail when the angle from which the measurements are taken is limited (Figure~\ref{fig:limited_angle_reconstruction}).
Recently, reconstruction algorithms based on deep learning produced promising results~\cite{mcleavy2021future}.
Adler and Öktem~\cite{adler2018learned} introduce a primal-dual method for tomographic reconstructing that relies on convolutional neural networks and is trained using algorithm unrolling. \C{Bubba et al.~\cite{bubba2019learning} combine an $\ell_1$-regularized shearlet solution with a neural network to predict insufficiently constrained shearlet coefficients.} Several methods combine the FBP algorithm with deep learning:
\sout{Würfl et al.~\cite{wurfl2018deep}  introduce a cone-beam back-projection layer that is used in a deep neural network to reconstruct images from limited angle data.} Dong et al.~\cite{dong2019deep} use a neural network to refine the sinogram obtained from the FBP reconstruction. The approach by Lee et al.~\cite{lee2018deep} interpolates a sparsely sampled sinogram using a neural network. Jin et al.~\cite{jin2017deep} reconstruct images from FBP reconstructions of limited angle data that exhibit strong artifacts.
\C{Other approaches, such as Schawb et al.~\cite{schwab2019deep} or Genzel et al.~\cite{genzel2022near}, use a data consistency term; the later in an iterative refinement scheme, winning the AAPM deep-learning sparse-view CT grand challenge~\cite{sidky2022report}.}
Generative adversarial networks~\cite{goodfellow2020generative} to reconstruct images from sinograms have been used by Yang et al.~\cite{yang2020tomographic}.
\C{Würfl et al.~\cite{wurfl2016deep, wurfl2018deep} learn compensation weights for a fixed back projection operator.
In addition to artifact removal in the image domain, Zhang et al.~\cite{zhang2020artifact} train a network in the projection domain.
Paschalis et al.~\cite{paschalis2004tomographic}, Zhu et al.~\cite{zhu2018image} and Yim et al.~\cite{yim2021limited} learn direct reconstruction on relatively small images and mention high computational costs, which arise from the use of fully connected layers. Adler et al.~\cite{adler2018learned} also note the difficulty of learning the full back projection operator.

\subsection{Our contribution}

Our contributions can be summarized as follows:
\begin{itemize}
\item In the context of the Helsinki Tomography Challenge, we demonstrate that fully learned backward operators are practical for images as large as $512 \times 512$ pixels and can achieve competitive performance, ultimately winning the first-place position in the challenge.
\item We achieve this via a carefully designed fully convolutional architecture, drawing inspiration from the recent ConvNeXt architecture \cite{liu2022convnet}.
\item The challenge poses the problem of varying starting angles of rotation. In previous approaches based on learned artifact removal, the problem was naturally solved by filtered back projection. Since we do not use filtered back projection, we propose the usage of a differentiable rotation layer instead.
\item Finally, we validate our approach with various ablation studies.
\end{itemize}\phantom{.}}

\subsection{Helsinki Tomography Challenge}
In our work, we consider the Helsinki Tomography Challenge 2022 (HTC 2022) open tomographic dataset \cite{meaney2022helsinki}, which was recorded at the Industrial Mathematics Computed Tomography Laboratory at the University of Helsinki.
The training data consists of five acrylic discs with 70mm diameter that had holes of different shapes. For each disk CT measurements of an angular range of $90\degree$ in $0.5 \degree$ increments were obtained, i.e., $\sinlen=181$ using a Hamamatsu Photonics C7942CA-22 X-ray image sensor.
The testing dataset consists of $21$ disks (each with different holes in them) and their corresponding CT measurements. The testing data was held back until all teams submitted their methods, however at the time of writing this paper we have access to the testing data as well, which allows us to thoroughly evaluate our method.

\begin{table}[h]
  \centering
  \caption{The levels of difficulty of the reconstruction task of the HTC 2022. The fewer measurements $\sinlen$ are given, the harder is the reconstruction task (left to right).}
  \label{tab:levels}
  \begin{tabular}{lccccccc}
    \toprule
    Level         & 1         & 2         & 3         & 4         & 5         & 6         & 7         \\
    \midrule
    $\sinlen$     & 181       & 161       & 141       & 121       & 101       & 81        & 61        \\
    Angular range & 90\degree & 80\degree & 70\degree & 60\degree & 50\degree & 40\degree & 30\degree \\
    \bottomrule
  \end{tabular}
\end{table}

The challenge asks to reconstruct images from a limited number of the full measurements. The number of consecutive angles determines the difficulty of the problem. Table \ref{tab:levels} gives an overview of the different levels.

\section{Generation of Synthetic Data}

\newcommand{\cw}{0.24}
\begin{figure}
  \centering
  \begin{subfigure}[t]{\cw\textwidth}
    \centering
    \includegraphics[height=2.5cm]{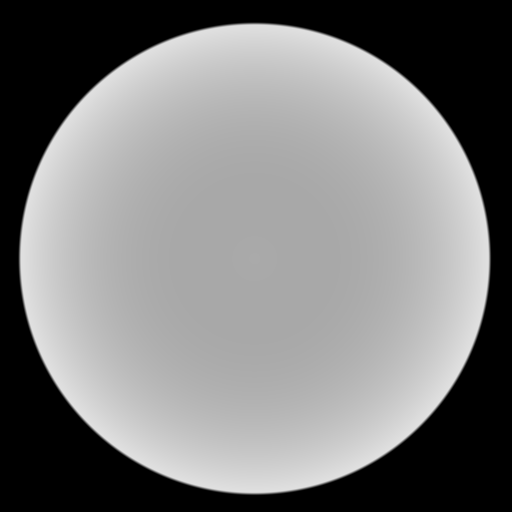}
    \captionsetup{justification=centering}
    \caption[(i)]{Empty disk.}
    \label{fig:emptydisk}
  \end{subfigure}
  \hfill
  \begin{subfigure}[t]{\cw\textwidth}
    \centering
    \newcommand{\imgwidth}{0.4cm}
    \newcommand{\nx}{0.85}
    \begin{tikzpicture}
      \node at (0, 0) {\includegraphics[width=\imgwidth]{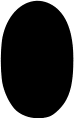}};
      \node at (1*\nx, 0) {\includegraphics[width=\imgwidth]{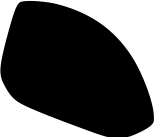}};
      \node at (2*\nx, 0) {\includegraphics[width=\imgwidth]{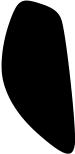}};
      \node at (0, -1*\nx) {\includegraphics[width=\imgwidth]{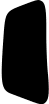}};
      \node at (1*\nx, -1*\nx) {\includegraphics[width=\imgwidth]{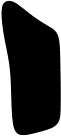}};
      \node at (2*\nx, -1*\nx) {\includegraphics[width=\imgwidth]{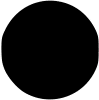}};
      \node at (0, -2*\nx) {\includegraphics[width=\imgwidth]{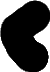}};
      \node at (1*\nx, -2*\nx) {\includegraphics[width=\imgwidth]{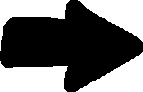}};
      \node at (2*\nx, -2*\nx) {\includegraphics[width=\imgwidth]{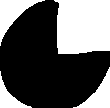}};
    \end{tikzpicture}
    \captionsetup{justification=centering}
    \caption{Extracted and handcrafted shapes.}
    \label{fig:shapes}
  \end{subfigure}
  \hfill
  \begin{subfigure}[t]{\cw\textwidth}
    \centering
    \newcommand{\imgwidth}{12mm}
    \newcommand{\nx}{1.3}
    \begin{tikzpicture}
      \node at (0, 0) {\includegraphics[width=\imgwidth]{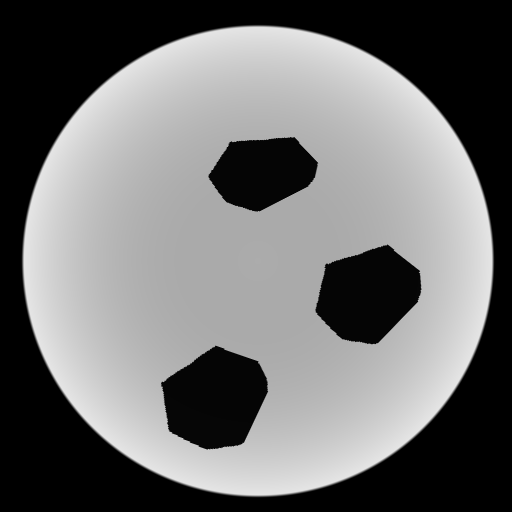}};
      \node at (1*\nx, 0) {\includegraphics[width=\imgwidth]{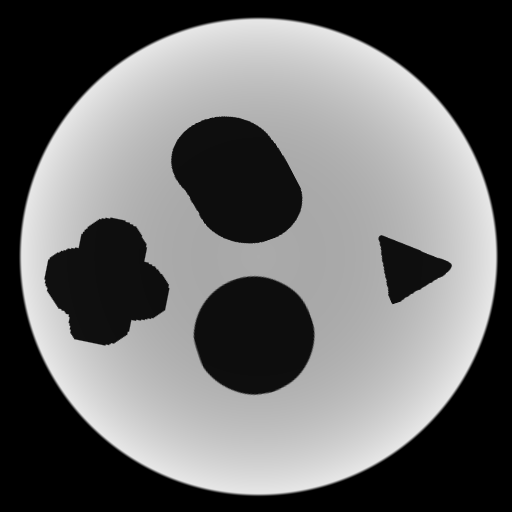}};
      \node at (0, -1*\nx) {\includegraphics[width=\imgwidth]{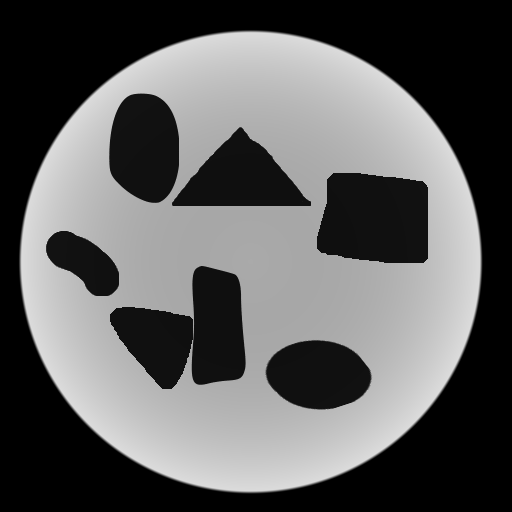}};
      \node at (1*\nx, -1*\nx) {\includegraphics[width=\imgwidth]{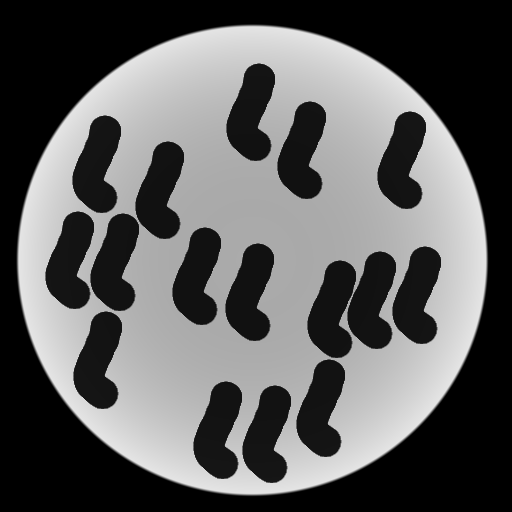}};
    \end{tikzpicture}
    \captionsetup{justification=centering}
    \caption{Disks filled with shapes.}
    \label{fig:shapesondisks}
  \end{subfigure}
  \hfill
  \begin{subfigure}[t]{\cw\textwidth}
    \centering
    \newcommand{\imgwidth}{12mm}
    \newcommand{\nx}{1.3}
    \begin{tikzpicture}
      \node at (0, 0) {\includegraphics[width=\imgwidth]{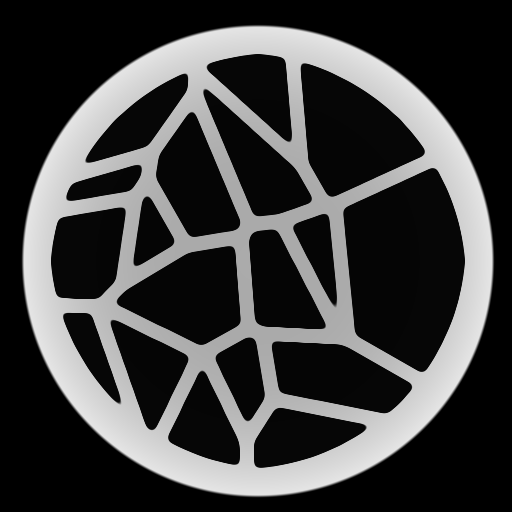}};
      \node at (1*\nx, 0) {\includegraphics[width=\imgwidth]{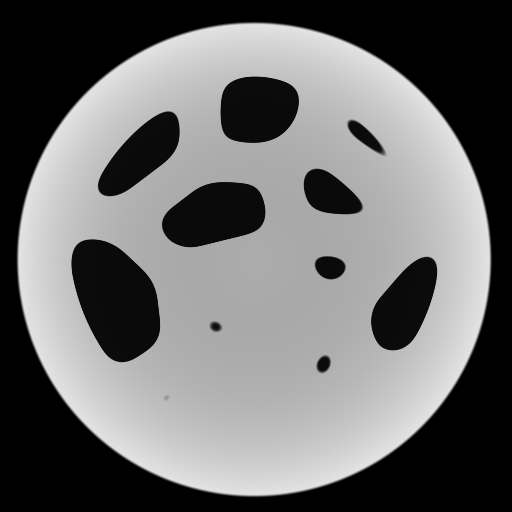}};
      \node at (0, -1*\nx) {\includegraphics[width=\imgwidth]{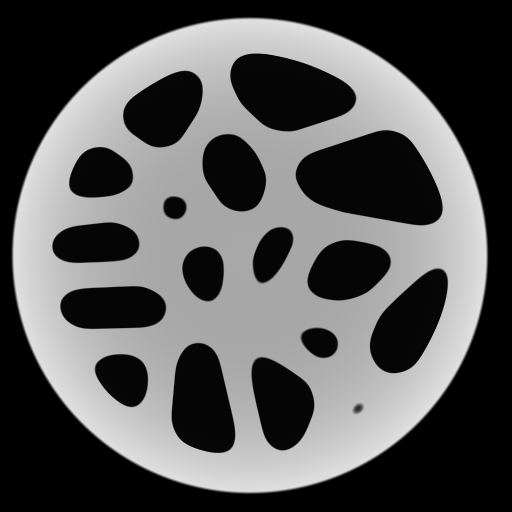}};
      \node at (1*\nx, -1*\nx) {\includegraphics[width=\imgwidth]{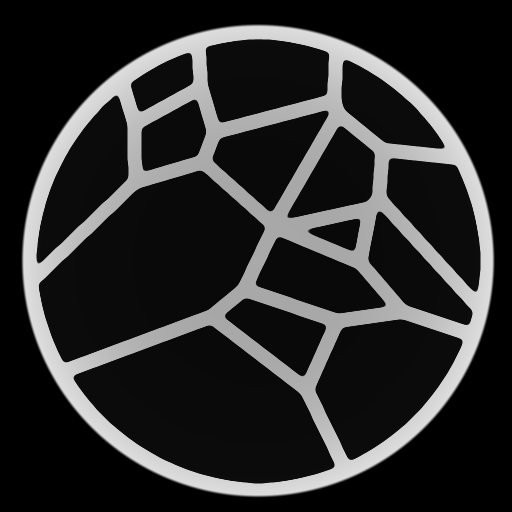}};
    \end{tikzpicture}
    \captionsetup{justification=centering}
    \caption{Disks filled with varied Voronoi cells.}
    \label{fig:voronoicells}
  \end{subfigure}
  \caption{An exemplary overview of the four steps that make up the generation process.}
  \label{fig:data-generation}
\end{figure}

The general idea of our approach is to train a neural network to predict images from subsampled sinograms. Since the dataset provided by the HTC 2022 organizers is too small to train a deep neural network (only five phantoms and their corresponding sinograms are given), we created a synthetic dataset of $200{,}000$~images that are visually similar to the provided phantoms. From these images, we calculate the corresponding sinograms for the CT problem.

\C{Synthetic data enjoys a long history in the context of computed tomography~\cite{shepp1974fourier} and limited angle CT~\cite{inouye1979image} to evaluate the efficacy of reconstruction methods. Recently, synthetic data has been used to train neural networks \cite{adler2018learned, barutcu2021limited, bubba2019learning, dong2019deep, goy2019high, jin2017deep, paschalis2004tomographic, sidky2022report}, for example circles, ellipses, or more problem-specific phantoms. It has been shown that synthetic data can suffice as a training set when evaluated on real data~\cite{douarre2018transfer, goy2019high}. We tailor our dataset specifically to the HTC 2022 challenge.}

\subsection{Phantom Simulation}
We generate the synthetic phantoms using the following steps, which we also illustrate in Figure~\ref{fig:data-generation}:
\begin{enumerate}[(A)]
  \item First, we create empty disks with varying position, radius and brightness (Figure~\ref{fig:emptydisk}). \C{The intention is that a neural network trained on this data might learn to perform well across the full range of varied parameters. We later verify that this assumption holds in Section~\ref{sec:positiona_invariance}.} We observe that the full angle reconstructions of the disks from the HTC 2022 dataset are darker in the center than at the edge. This can be attributed to an effect called \textit{beam hardening} due to the interaction of X-rays with materials of different density \cite{brooks1976beam}. In our case, the objects are made from acrylic and are surrounded by air.
        We model the brightness of the disk as a third degree polynomial with coefficients $c_i$ that is parameterized by the distance $d$ to the center of the disk
        \begin{align*}
          \brightness(d) = \sum_{i=0}^3 c_i d^i.
        \end{align*}
        We fit the coefficients of the polynomial using the disk provided and vary the coefficients slightly to increase dataset diversity.

        Furthermore, we observe that the edges of the disk are smooth, i.e., they do not transition instantly from white to black, but rather fade out over several pixels. We approximate this transition using Hermite interpolation using the smoothstep function \begin{align*}
          \smoothstep(d, e_1, e_2) = \begin{cases}
                                       0,                                                                                       & \text{if } d\in(-\infty, e_1] \\
                                       3\left(\dfrac{d - e_1}{e_2 - e_1}\right)^2 - 2\left(\dfrac{d - e_1}{e_2 - e_1}\right)^3, & \text{if }  d\in(e_1, e_2)    \\
                                       1,                                                                                       & \text{if }  d\in[e_1, \infty)
                                     \end{cases}
        \end{align*}
        where $e_1$ and $e_2$ denote the lower and upper edge, respectively. This smoothstep function is also commonly used in the field of computer graphics~\cite{texturing, opengl}.
  \item Next, we extract coherent components from the given training phantoms and then extend the resulting set to include additional handcrafted shapes in order to increase the diversity of our dataset (Figure~\ref{fig:shapes}). Since the test dataset was unknown to us at this point, we made a guess as to which shapes might appear in the test dataset. We favored shapes with rounded corners because we believed it to be more difficult to cut out pointy shapes from acrylic. In addition, the training data exclusively contained shapes with rounded corners.
  \item We sequentially place shapes on the first $80\%$ of the disks generated in step (A), making sure that the shapes placed on the same disk do not overlap (Figure~\ref{fig:shapesondisks}). For each disk, we vary the number of shapes that we place on the disk as well as the minimal distance to other shapes and the edge of the disk.
        Empty disks are also a possibility in the case where the number of shapes to be placed on a disk is zero.
        For some disks, we also randomly vary the scale and rotation of the shapes and place different types of shapes on the disk.
  \item We fill the other $20\%$ of the disks from step (A) with smoothed Voronoi cells with varying number and positions of center points, degree of corner smoothing and border radius (Figure~\ref{fig:voronoicells}).
\end{enumerate}

\subsection{Sinogram Simulation}

The previously generated phantom images can now be used to compute synthetic sinograms by multiplying the system matrix with the vectorized image (Equation~\ref{eq:problem}).
To define the \sout{linear system}\C{operator matrix,} we follow the geometric definition given by the organizers of the Helsinki Tomography Challenge 2022 \cite{meaney2022helsinki} as illustrated in Figure~\ref{fig:setup}.
Specifically, we utilized the same number of detector pixels, physical pixel size, distance from the X-ray source to the center of the target and distance from the center to the detector to create a 2D flat fan projector using the ASTRA Toolbox \cite{van2016fast, van2015astra}.
In this way, we acquire sinograms with an angular range of $360$ degrees and an angular increment of $0.5$ degrees, resulting in $721$ sinogram rows that can be subsampled at will to define the limited angle CT problem of interest.

\section{Method}

\label{sec:method}

To solve the inverse problem, we learn a deep neural network~$f_\theta$ with
parameters~$\theta$ in an end-to-end manner. The network is trained to minimize
the mean squared error between the true images $x$ and its predictions
$f_\theta(y_{\alpha:\beta})$, given limited angle sinograms $y_{\alpha:\beta}$
with angles $\alpha, \beta \in \{0,0.5,\ldots,360\}$ such that $30 \leq \beta -
  \alpha \leq 90$. The starting angle $\alpha$ of the sinogram determines the
rotation of the image. We facilitate the learning task by rotating the output of
the neural network by $\alpha$, so that this transformation does not have to be
learned. Thus, we define our learning objective as
\begin{equation}
  \min_\theta\, \mathbb{E}_{(x,y),(\alpha,\beta)}\!\Big[ \| R_\alpha( f_\theta(y_{\alpha:\beta}) ) - x\|_2^2 \Big]\hspace{-1pt},
  \label{eq:loss}
\end{equation}
where $R_\alpha(\cdot)$ denotes a counterclockwise rotation through angle
$\alpha$ about the origin. The $(x,y)$-pairs are sampled jointly from the
dataset and the angular ranges $\beta - \alpha$ are sampled from $30$ to $90$ degrees in $10$ degree steps with descending probabilities to give a larger weight to more difficult reconstructions from fewer angles.
The starting angle $\alpha$ is sampled from a uniform distribution such that the partial sinogram $y_{\alpha:\beta}$ does not wrap around, i.e., $\beta \le 360^{\degree}$.

\begin{figure}
  \resizebox{\linewidth}{!}{\input{figures/network.tex}}
  \caption{Illustration of our neural network architecture.}
  \label{fig:architecture}
\end{figure}

The motivation behind our neural network architecture is to create a bottleneck:
several convolutional layers reduce the spatial dimensions of the sinogram from
$181 \times 560$ pixels down to a representation of size $8 \times 8 \times
  512$. This is followed by multiple transposed convolutions to obtain an image
prediction of size $512 \times 512$. The representation should contain abstract
information about the sinogram, and the convolutional layers that operate in
this low resolution can cover and process large areas of the original input.

\subsection{Implementation Details.}

Figure \ref{fig:architecture} shows our proposed neural network. We
utilize residual blocks with skip connections~\cite{he2016deep} \C{to facilitate gradient flow for backpropagation,
but do not use long-range skip connections as in the U-Net~\cite{ronneberger2015u} architecture
commonly employed in methods that aim to remove artifacts introduced by filtered back projection.
For the latter, a strong correspondence between features in the encoder and decoder part of the network is desirable.
However, a pixel in the reconstruction can potentially be influenced by any pixel in the sinogram, depending on projection geometry, which limits the usefulness of long-range skip connections, since they only transport information for spatially aligned pixels.

To achieve top scores in a challenge, a fast turnaround time is crucial to evaluate changes to the dataset composition or general method. An efficient network is therefore desirable. An important aspect to reach that goal are large initial strides in the early convolutional layers to quickly decrease the size of feature maps, leading to more efficient computation.}
The authors of ConvNeXt \cite{liu2022convnet} showed many improvements to the standard ResNet
architecture. We found that some of these changes are also beneficial for our
network:
\begin{enumerate}
  \item Inverted bottlenecks: The number of hidden channels inside residual
        blocks is larger than the input channels (instead of reducing the number
        of channels).
  \item We increase the kernel size in residual blocks from $3 \times 3$ to $5
          \times 5$ \C{to compensate for the large initial strides.}
  \item We use a single GELU activation \cite{hendrycks2016gaussian} per
        residual block instead of two RELU activations per block.
  \item Separate downsampling layers: Convolutional layers with stride 2 reduce
        the spatial dimensions only outside of the residual blocks, followed by
        batch normalization layers \cite{ioffe2015batch}.
\end{enumerate}

\noindent
We minimize the mean squared error in Equation~\ref{eq:loss} with the Adam optimizer~\cite{kingma2014adam} using the default hyperparameters and a learning rate of
$10^{-4}$. We use the Kornia library~\cite{riba2020kornia} to compute the
rotation $R_\alpha(\cdot)$, as it needs to be differentiable.

Our neural network supports limited angular ranges between $30$ and $90$
degrees. The architecture expects the same input shape for any angular range,
therefore the remaining rows of sinograms with less than $90$ degrees are padded
with zeros. We explicitly specify the angular range of the input by
concatenating a binary mask on the channel dimension. In theory, this could help the neural
network differentiate between black pixels and pixels without information. \C{Further experiments revealed that the same approach without the concatenated binary mask achieves the same performance.}

Our implementation is available at \href{https://github.com/99991/HTC2022-TUD-HHU-version-1}{https://github.com/99991/HTC2022-TUD-HHU-version-1}.

\section{Experimental Evaluation}

We evaluate our method on the dataset of the Helsinki Tomography Challenge 2022 \cite{meaney2022helsinki}. The test dataset consists of seven increasingly more difficult levels, encompassing three images each.

The official score for the challenge was computed as the sum of Matthews Correlation Coefficient (MCC) over the three thresholded binary limited angle reconstructions for each individual level. Each team had to pass a certain score in order to advance to the next level. Finally, the teams were ranked by their score on level~7~(Figure~\ref{fig:colorfulresults}). A perfect score of $3$ would indicate that the reconstructed binary masks are identical to the full angle reconstructions, while a score of $-3$ indicates complete disagreement. We use simple mean thresholding to generate binary masks from our raw predictions.

In addition, we evaluate the peak signal-to-noise ratio (PSNR) and the structural similarity index measure (SSIM) of the raw reconstructions of our method before thresholding. We compare it to the filtered back projection method (FBP) \C{and a modification of the method by Jin et al.~\cite{jin2017deep}, who refine a reconstruction obtained by FBP with a U-Net. For easier comparison to our fully learned method, we replace the U-Net with our neural network architecture, which we adapt to inputs of size $512 \times 512$ by adjusting the strides of the encoder's convolutional layers. We also re-introduce long-range skip connections where applicable.}

\section{Results}

\begin{figure}
  \centering
  \hspace{-14mm}
  \includegraphics[width=\figurewidth]{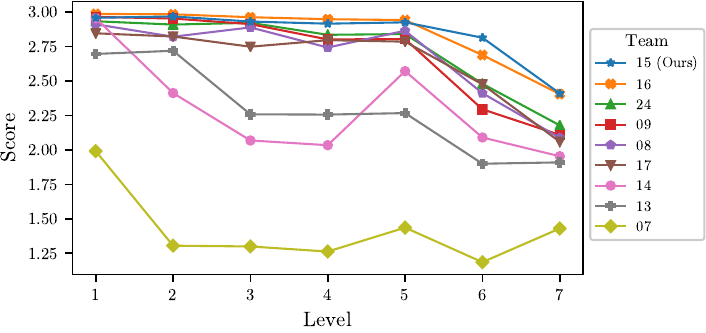}
  \caption{Comparison of Matthews Correlation Coefficient (MCC) for different teams participating in the Helsinki Tomography Challenge 2022. The score is the sum over three different reconstructions per difficulty level. The scores of our method (Team 15) are denoted by the blue line with star-shaped markers.}
  \label{fig:colorfulresults}
\end{figure}

Figure~\ref{fig:colorfulresults} shows the scores of the different teams which participated in the Helsinki Tomography Challenge 2022 for seven difficulty levels. As can be seen, our method (Team 15) did not initially lead the field during the easier difficulty levels, presumably because we only trained a single neural network that is applied to all difficulty levels, while other teams choose to train individually tuned networks for the different levels. However, as the challenge progresses, our method overtakes the second-best submission at level 6 and is able to win by a small margin at level 7, \C{followed by Team 16\footnote{\url{https://github.com/alexdenker/htc2022_LPD}}, who employ a learned primal-dual method~\cite{adler2018learned}, also making use of synthetically generated training data. The best non-learning-based approach by Team 24\footnote{\url{https://github.com/TomographicImaging/CIL-HTC2022-Algo2}} achieved rank 3 with a carefully regularized least-squares approach.}

\begin{table}
    \centering
    \begin{tabular}{cccccccccc}
        \toprule
        Level & \multicolumn{3}{c}{MCC} & \multicolumn{3}{c}{PSNR}       & \multicolumn{3}{c}{SSIM}                                                                                             \\
        \midrule \\
        & \multicolumn{1}{c}{\rotatebox[origin=l]{90}{FBP}} & \multicolumn{1}{c}{\rotatebox[origin=l]{90}{FBP+NN}} & \multicolumn{1}{c}{\rotatebox[origin=l]{90}{Our method}} & \multicolumn{1}{c}{\rotatebox[origin=l]{90}{FBP}}  & \multicolumn{1}{c}{\rotatebox[origin=l]{90}{FBP+NN}} & \multicolumn{1}{c}{\rotatebox[origin=l]{90}{Our method}} & \multicolumn{1}{c}{\rotatebox[origin=l]{90}{FBP}} & \multicolumn{1}{c}{\rotatebox[origin=l]{90}{FBP+NN}}& \multicolumn{1}{c}{\rotatebox[origin=l]{90}{Our method}} \\
        \cmidrule(lr){2-4}  \cmidrule(lr){5-7} \cmidrule(lr){8-10}
        1 & 1.92 & 2.92 & 2.96 & 10.74 & 24.34 & 26.01 & 0.46 & 0.66 & 0.67  \\
        2 & 2.00 & 2.96 & 2.97 & 10.54 & 23.43 & 25.61 & 0.45 & 0.62 & 0.68  \\
        3 & 1.89 & 2.94 & 2.93 & 10.48 & 22.70 & 23.34 & 0.42 & 0.69 & 0.69  \\
        4 & 1.83 & 2.92 & 2.92 & 10.33 & 21.65 & 23.36 & 0.39 & 0.70 & 0.70  \\
        5 & 1.54 & 2.89 & 2.93 &  9.86 & 20.88 & 23.94 & 0.38 & 0.66 & 0.68  \\
        6 & 1.07 & 2.45 & 2.81 &  8.93 & 15.36 & 21.17 & 0.34 & 0.61 & 0.68  \\
        7 & 0.84 & 2.31 & 2.41 &  7.89 & 14.44 & 16.40 & 0.31 & 0.59 & 0.63  \\
        \bottomrule
    \end{tabular}
    \caption{Quantitative evaluation on the HTC 2022 test data with 3 images per level. PSNR and SSIM are calculated on the non-binary reconstructions. The reported MCC value is calculated as the sum of the MCC values of the binary reconstructions.}
    \label{tab:results}
\end{table}

\begin{figure}
  \centering
  \setlength{\tabcolsep}{5pt}
  \newcommand{\imgwidth}{2.3cm}
  \newcommand{\impad}{-2mm}
  \renewcommand{\arraystretch}{2}
  \begin{tabular}{rcccc}
    Level & FBP & \C{FBP + NN} & Our method & Ground truth \\
    1  ($90\degree$) &
    \includegraphics[width=\imgwidth, align=c]{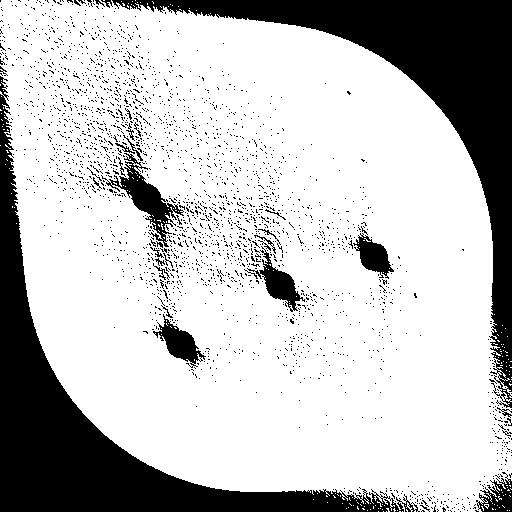}   \hspace{\impad} &
    \includegraphics[width=\imgwidth, align=c]{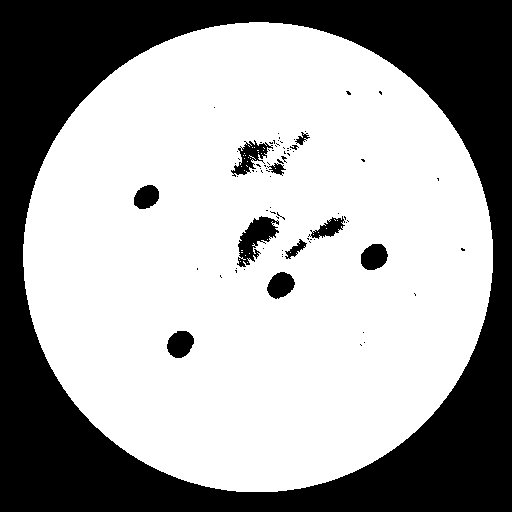}       \hspace{\impad} &
    \includegraphics[width=\imgwidth, align=c]{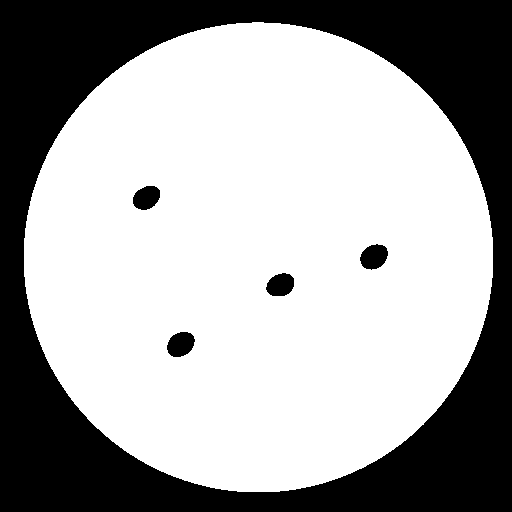}       \hspace{\impad} &
    \includegraphics[width=\imgwidth, align=c]{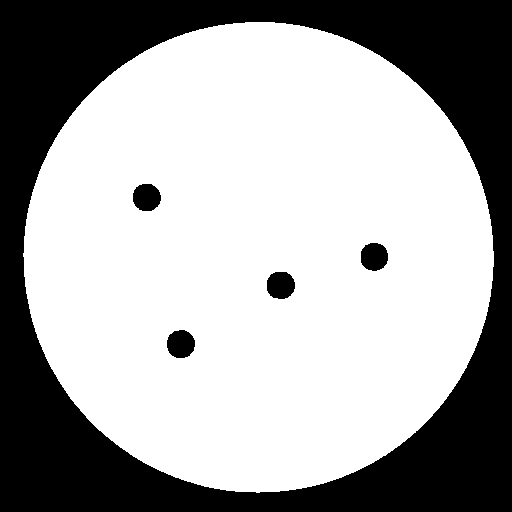} \vspace{0.25cm} \\
    2 ($80\degree$)  &
    \includegraphics[width=\imgwidth, align=c]{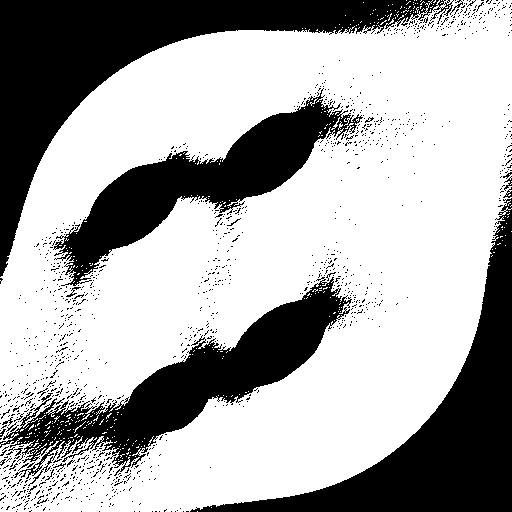}   \hspace{\impad} &
    \includegraphics[width=\imgwidth, align=c]{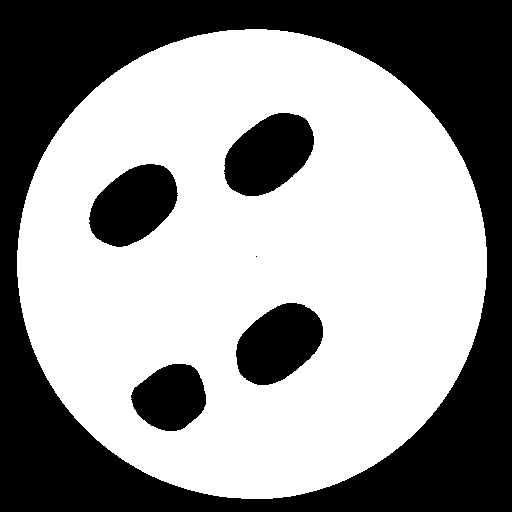}       \hspace{\impad} &
    \includegraphics[width=\imgwidth, align=c]{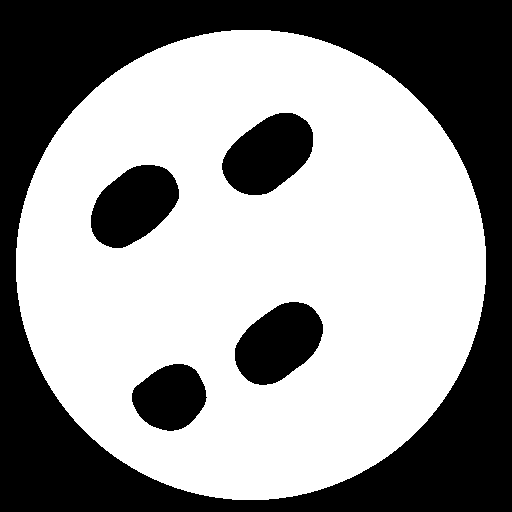}       \hspace{\impad} &
    \includegraphics[width=\imgwidth, align=c]{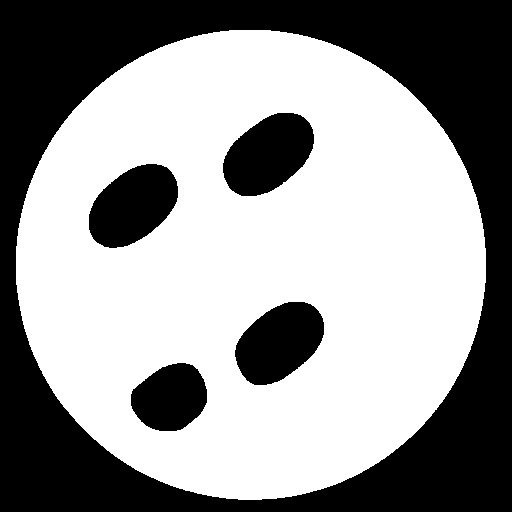} \vspace{0.25cm} \\
    3 ($70\degree$)  &
    \includegraphics[width=\imgwidth, align=c]{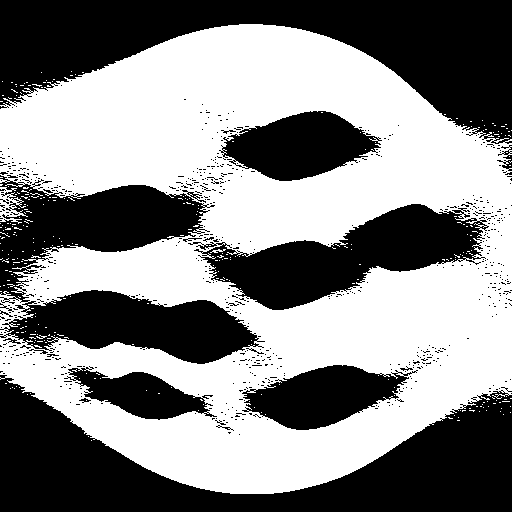}   \hspace{\impad} &
    \includegraphics[width=\imgwidth, align=c]{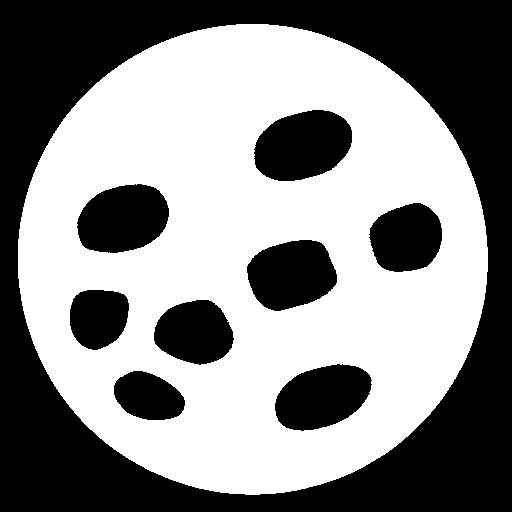}       \hspace{\impad} &
    \includegraphics[width=\imgwidth, align=c]{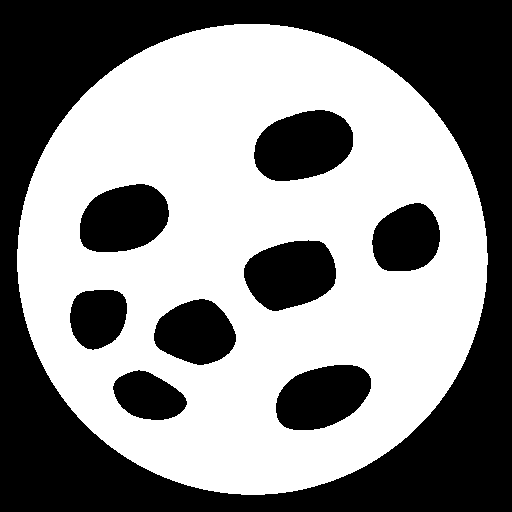}       \hspace{\impad} &
    \includegraphics[width=\imgwidth, align=c]{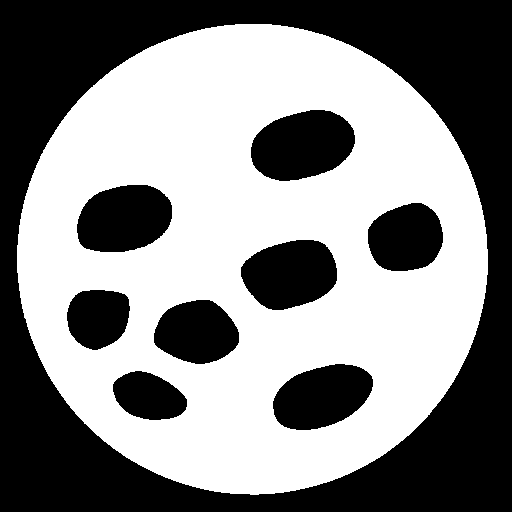} \vspace{0.25cm} \\
    4 ($60\degree$)  &
    \includegraphics[width=\imgwidth, align=c]{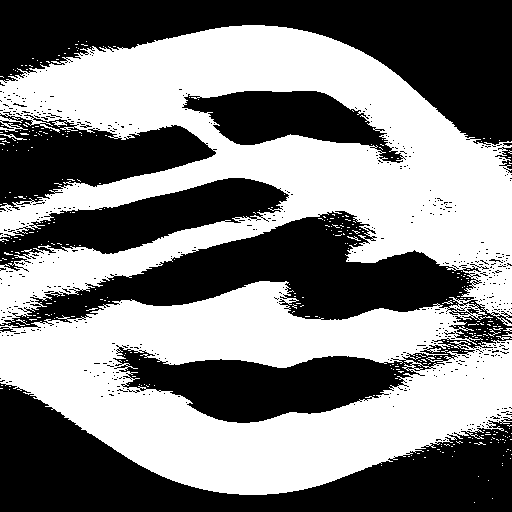}   \hspace{\impad} &
    \includegraphics[width=\imgwidth, align=c]{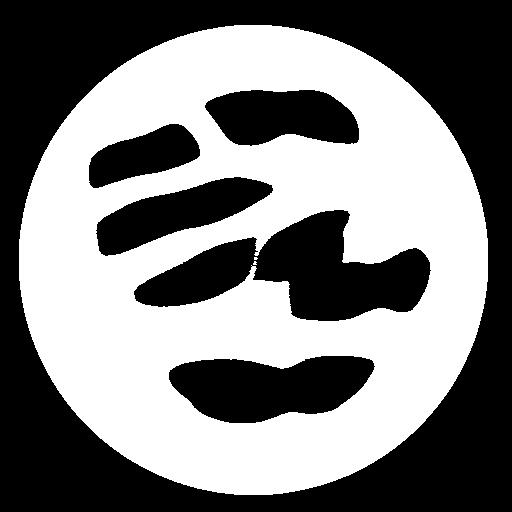}       \hspace{\impad} &
    \includegraphics[width=\imgwidth, align=c]{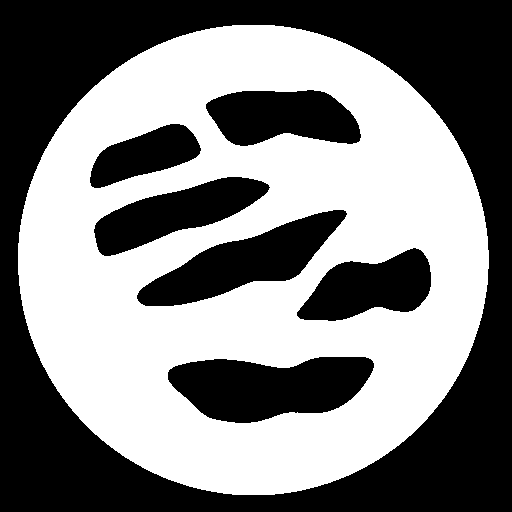}       \hspace{\impad} &
    \includegraphics[width=\imgwidth, align=c]{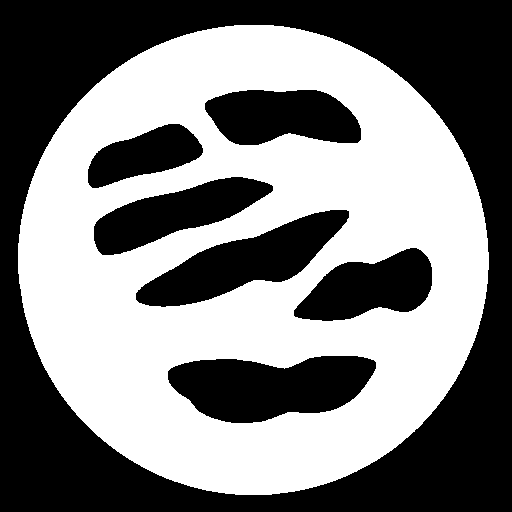} \vspace{0.25cm} \\
    5 ($50\degree$)  &
    \includegraphics[width=\imgwidth, align=c]{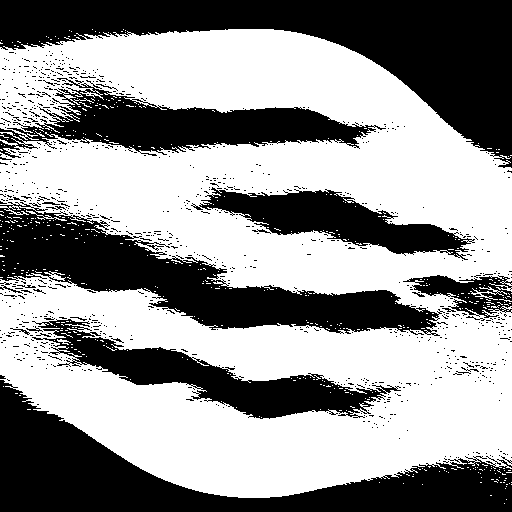}   \hspace{\impad} &
    \includegraphics[width=\imgwidth, align=c]{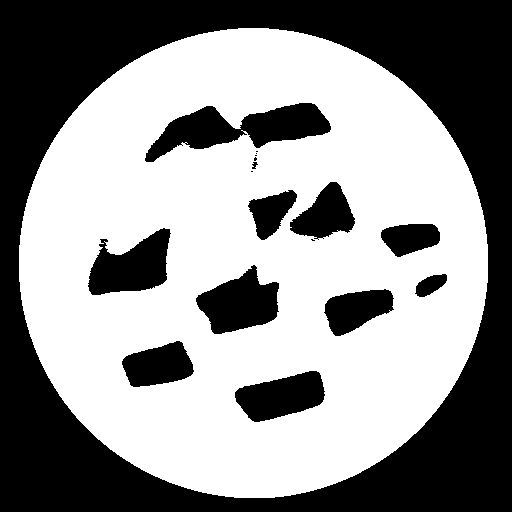}       \hspace{\impad} &
    \includegraphics[width=\imgwidth, align=c]{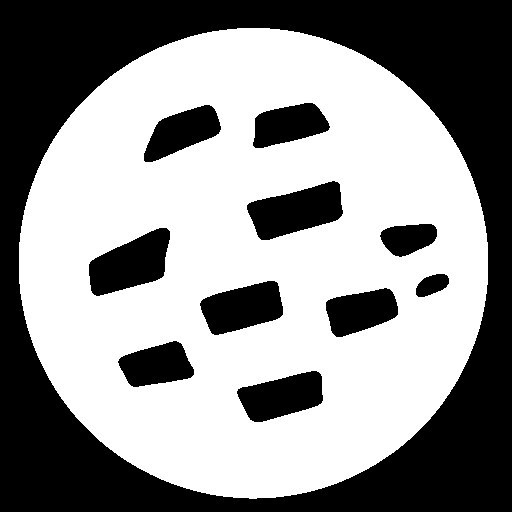}       \hspace{\impad} &
    \includegraphics[width=\imgwidth, align=c]{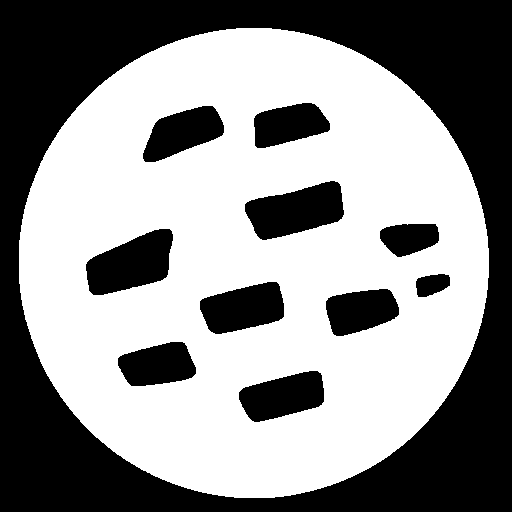} \vspace{0.25cm} \\
    6 ($40\degree$)  &
    \includegraphics[width=\imgwidth, align=c]{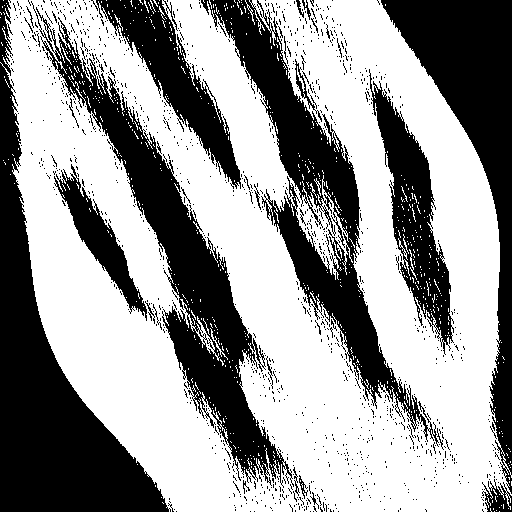}   \hspace{\impad} &
    \includegraphics[width=\imgwidth, align=c]{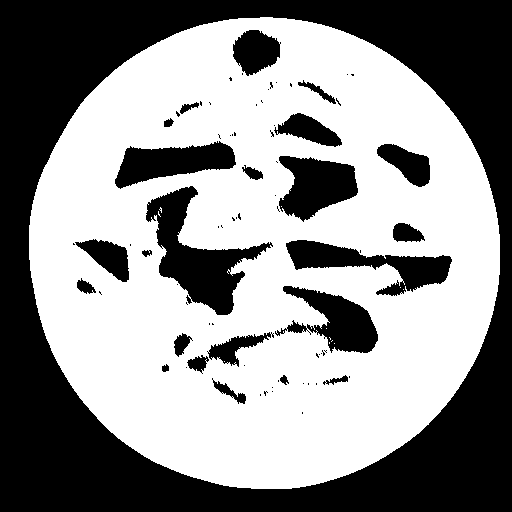}       \hspace{\impad} &
    \includegraphics[width=\imgwidth, align=c]{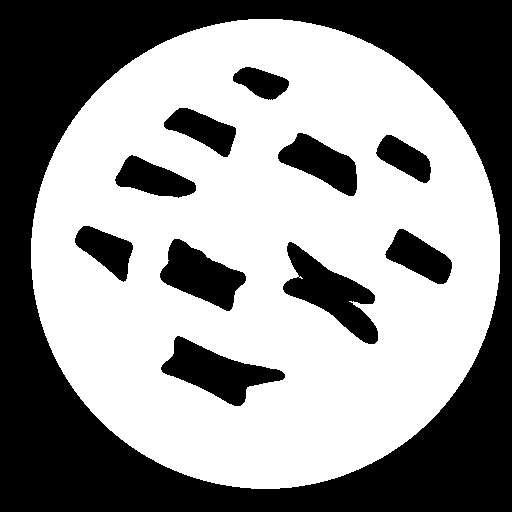}       \hspace{\impad} &
    \includegraphics[width=\imgwidth, align=c]{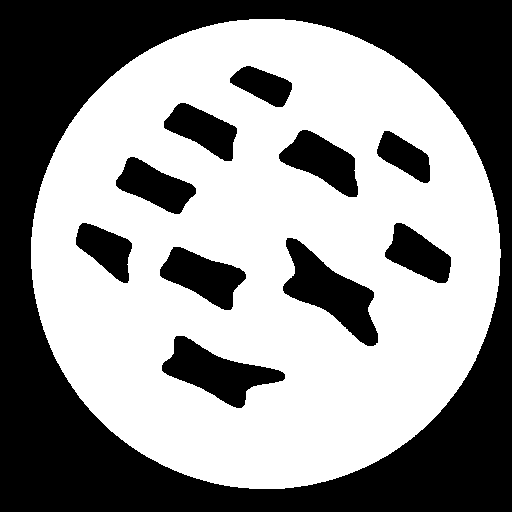} \vspace{0.25cm} \\
    7 ($30\degree$)  &
    \includegraphics[width=\imgwidth, align=c]{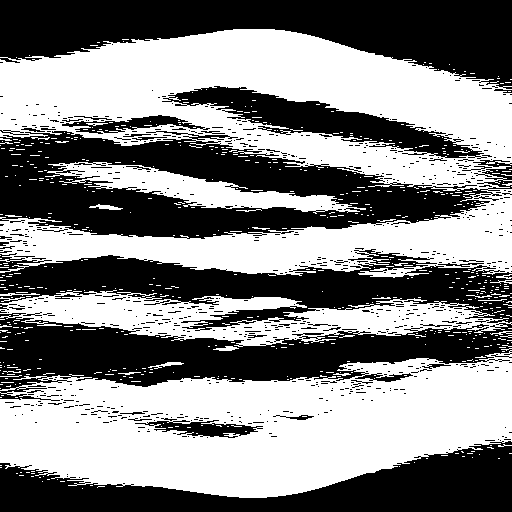}   \hspace{\impad} &
    \includegraphics[width=\imgwidth, align=c]{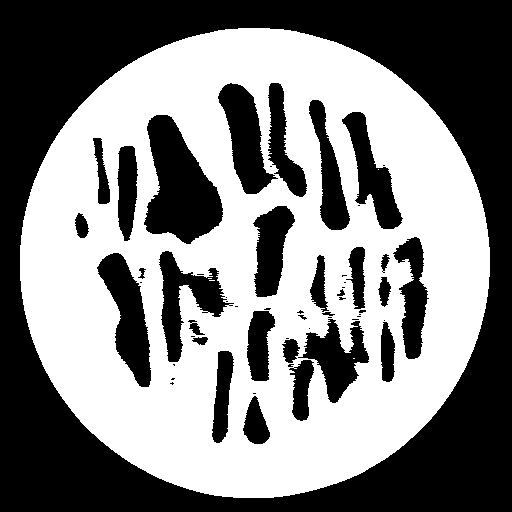}       \hspace{\impad} &
    \includegraphics[width=\imgwidth, align=c]{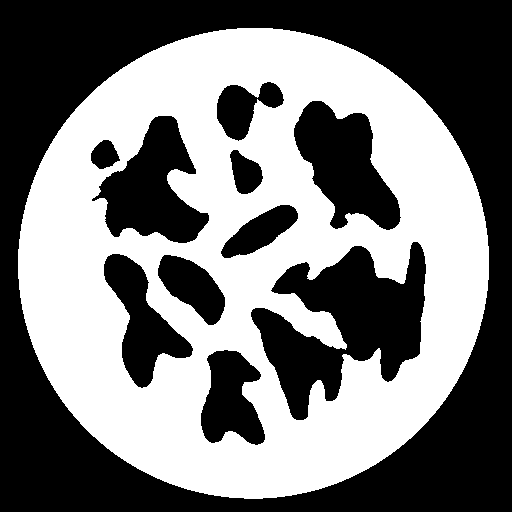}       \hspace{\impad} &
    \includegraphics[width=\imgwidth, align=c]{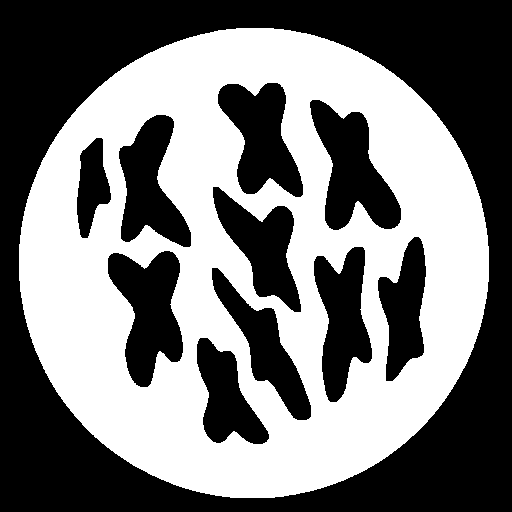} \vspace{0.25cm} \\
  \end{tabular}
  \caption{Qualitative comparison of binarized reconstructions from limited angle sinograms.}\label{fig:reconstructions}
\end{figure}

Our method compares favorably to the filtered back projection method, beating it with respect to PSNR, SSIM and MCC on every level (Table~\ref{tab:results}). \C{Refining the result of the filtered back projection with our modified U-Net drastically improves reconstruction quality, but still falls short of our fully learned method. We observe that the U-Net is generally able to suppress most artifacts, but sometimes replaces them with various shapes seen in the training dataset. Reconstructions are shown in Figure~\ref{fig:reconstructions}.}

\subsection{Runtime Performance}

Once the network is trained, our method does not require slow iterative optimization. A single forward pass through the network is sufficient to compute the reconstruction for a new sinogram, which makes our method exceptionally fast. Reconstructing one individual image takes only 13 milliseconds on a GeForce RTX 3060 Mobile GPU with 6 GB of VRAM. Batch processing of images decreases the runtime even more to as little as 8 milliseconds per image for a batch size of 10.

\subsection{Ablation Studies}

We examine the impact of various changes to our training procedure and to our synthetically generated dataset. For all ablation studies, we train the same neural network architecture for 100 epochs and evaluate the MCC score on the test split of the HTC 2022 dataset.

\subsubsection{Training one model per difficulty level} 
\begin{figure}
  \centering
  \includegraphics[width=\figurewidth]{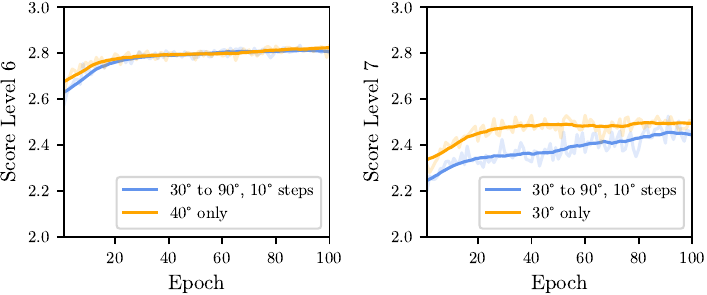}
  \caption{MCC score during training of a model over 100 epochs on sinograms with an angular range of 40 degrees (left, orange) and 30 degrees (right, orange) compared to our baseline trained on multiple angular ranges (blue).}
  \label{fig:3040deg}
\end{figure}

In our official submission to the Helsinki Tomography Challenge, we trained a single model to reconstruct images from sinograms with multiple different angular ranges. While this has the advantage that only a single model has to be trained, it is conceivable that a model trained on sinograms with a fixed angular range might produce superior reconstructions. To this end, we train one model that reconstructs images exclusively from sinograms with an angle of 30 degrees, which corresponds to difficulty level 7 of the Helsinki Tomography Challenge. In addition, we train a second model for difficulty level 6 with an angular range of 40 degrees. We compare to a baseline model trained on sinograms with 30, 40, 50, 60, 70, 80 as well as 90 degrees. The evolution of the MCC score on the test split of the HTC 2022 dataset over a training run of 100 epochs can be seen in Figure~\ref{fig:3040deg}.

The scores for the model trained on level 6 do not deviate greatly from the baseline model, mainly because both models already produce highly accurate reconstructions, which leaves little room for improvement. However, the model trained entirely on sinograms of level 7 outperforms the baseline. This suggests that it is beneficial to train individual models for exceedingly difficult tasks, while it could be economical to train a single model to simultaneously solve several easier tasks.

\subsubsection{Generalization with respect to the angular range of sinograms} 
\begin{figure}
  \centering
  \includegraphics[width=\figurewidth]{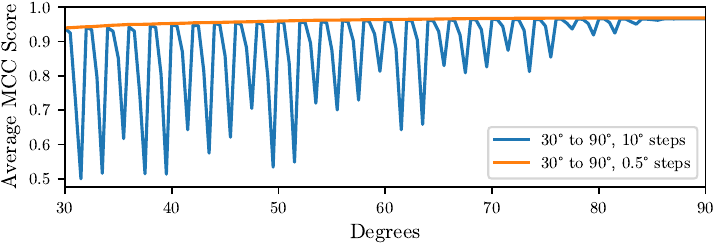}
  \caption{Test scores for two models trained on sinograms with different numbers of angular ranges.}
  \label{fig:angular_range_generalization}
\end{figure}

One might ask the question whether it is necessary to train a neural network on sinograms of every angular range, or if good test performance can be achieved by training on fewer angles. To test this, we train a baseline model on sinograms with an angular range as described in Section~\ref{sec:method} and compare it to another model trained on angular ranges sampled uniformly from 30 to 90 degrees with 0.5 degree increments. We evaluate the model performance by truncating the 21 sinograms of the test split of the HTC 2022 dataset to a given angular range and computing the average MCC score (Figure~\ref{fig:angular_range_generalization}).

Three main observations can be made. First, both models exhibit similar performance when only considering angular ranges from 30 to 90 degrees with 10 degree increments. Second, the model trained on these angular ranges does not generalize to all angular ranges in between. Third, the model still performs well on sinograms with an angular range of multiples of 2 degrees. This is an artifact of the model architecture consisting of an initial convolutional layer with kernel size and stride 4, which corresponds to 2 degrees since the sinograms have a resolution of $0.5$ degrees. It can be concluded that, if a model should be tested on angular ranges of any size, it is important to integrate that assumption in the training procedure.

\begin{figure}[b]
  \centering
  \includegraphics[width=\figurewidth]{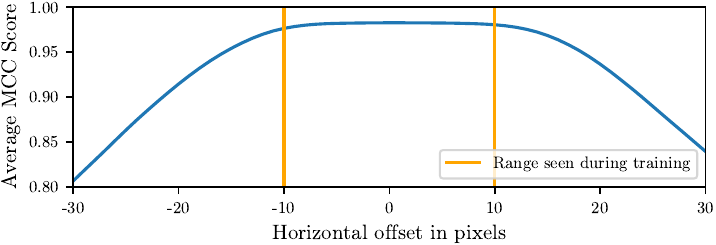}
  \caption{\C{Averaged MCC scores for 100 synthetically generated images translated horizontally by up to 30 pixels.}}
  \label{fig:positional_invariance}
\end{figure}

\C{\subsubsection{Positional invariance} \label{sec:positiona_invariance} When generating the synthetic phantom images, we varied their spatial position to promote spatial invariance during reconstruction. We quantitatively verify the success of this augmentation by computing the average MCC score (Figure~\ref{fig:positional_invariance}) for 100 horizontally translated images, which have been reconstructed from 30 degree limited angle sinograms. From the given training data and the challenge description, we estimated that the displacement of the test images, which were secret at the time of training, is not likely to vary by more than 10 pixels, determining our training parameters.

It can be seen that the reconstruction is accurate as long as the translation is not too large. As the offset increases beyond the range seen during training, the performance continually decreases.}

\subsubsection{Extending the training dataset with additional shapes} 
\begin{figure}
  \centering
  \begin{minipage}{.345\textwidth}
    \centering
    \includegraphics[width=1.0\linewidth]{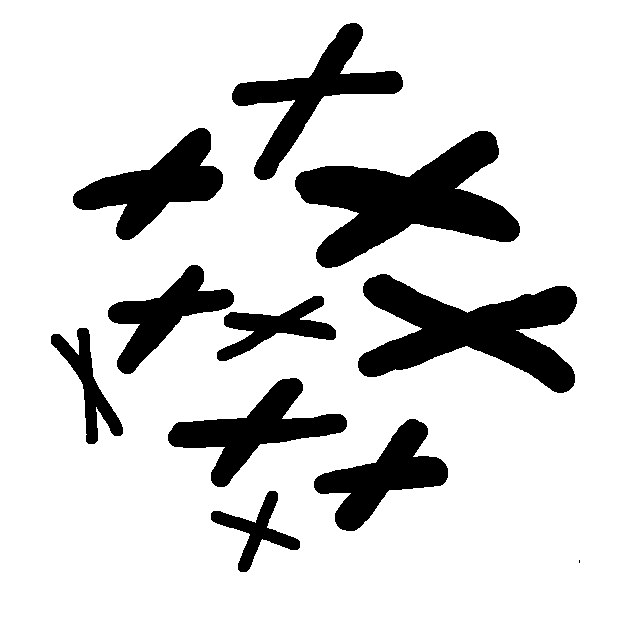}
    \captionsetup{width=0.847\linewidth}
    \captionof{figure}{Additional cross shapes.}
    \label{fig:crosses}
  \end{minipage}\hspace{0mm}
  \begin{minipage}{.63\textwidth}
    \centering
    \includegraphics[width=\linewidth]{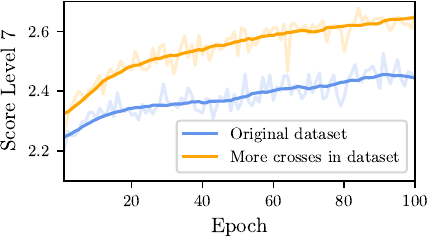}
    \captionsetup{width=0.9\linewidth}
    \captionof{figure}{MCC scores for model trained on dataset with additional cross shapes.}
    \label{fig:with_crosses}
  \end{minipage}
\end{figure}

For difficulty level 7 of the HTC 2022 dataset, all holes in the test disks are cross-shaped. Since our synthetic dataset contains relatively few concave shapes, one might wonder whether test performance could be improved by introducing additional cross-shaped objects into our dataset. We note that this experiment could only be conducted in hindsight, since the type of shapes in the test dataset was unknown before the challenge concluded. For a fair evaluation with respect to the HTC 2022, it would be improper to look at the test dataset. However, for practical applications, the type of shape to be scanned may be known in advance, even if the exact position, scale or rotation is unknown. To investigate the performance on this new problem formulation, we enrich our dataset of 108 shapes with 10 additional hand-drawn crosses (Figure~\ref{fig:crosses}) and compare the MCC scores (Figure~\ref{fig:with_crosses}).

It is apparent that the reconstruction accuracy increases significantly, which highlights the importance of a close match between training and test dataset. In addition, it can be seen that the test score still has not fully converged after 100 epochs of training, which would justify further training or an even higher number of cross-shaped objects in the training dataset.

\ifthenelse{\boolean{iscleanversion}}{}{
\subsubsection{\sout{Importance of sinogram validity mask}} \sout{We train a neural network which is capable of reconstructing images from sinograms with an angular range from 30 up to 90 degrees with $0.5$ degree steps. For technical reasons, it is preferable that the input shape of the sinogram is always the same. Therefore, we pad sinograms with fewer than 181 rows with zeros and provide an additional sinogram validity mask to the neural network to disambiguate missing measurements from zero measurements. In Figure~\ref{fig:sinogrammask}, we compare the MCC score on difficulty level 7 of the HTC 2022 dataset for neural networks trained with and without the sinogram validity mask. It appears that the difference between scores is negligible for this dataset.}

\begin{figure}
  \centering
  \includegraphics[width=\figurewidth]{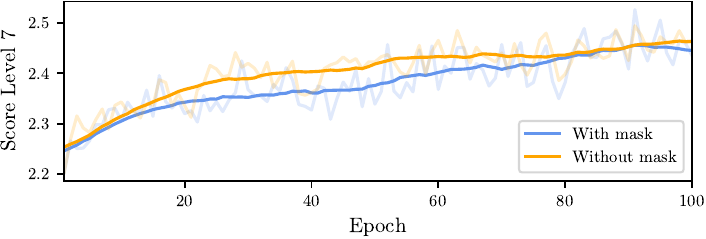}
  \caption{\sout{Comparison of MCC score with and without sinogram validity mask.}}
  \label{fig:sinogrammask}
\end{figure}
}

\subsubsection{Dataset size} 
\begin{figure}
  \centering
  \includegraphics[width=\figurewidth]{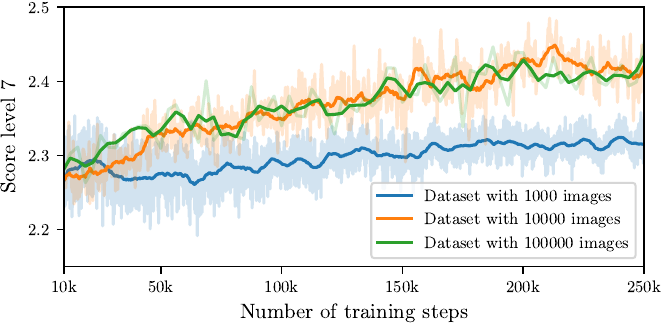}
  \caption{Comparison of MCC scores for models trained on datasets with different size on difficulty level 7 of the HTC 2022.}
  \label{fig:dataset_size}
\end{figure}

We investigate the effect of smaller dataset sizes on the performance of our model (Figure~\ref{fig:dataset_size}). For a fair comparison, we evaluate the MCC score after the same number of parameter updates for each model. This means that the models are trained for a different number of epochs as the dataset sizes differ.

\subsection{Limitations}
\begin{figure}
    \centering
    \begin{subfigure}[t]{.25\textwidth}
        \centering
        \includegraphics[height=2.5cm]{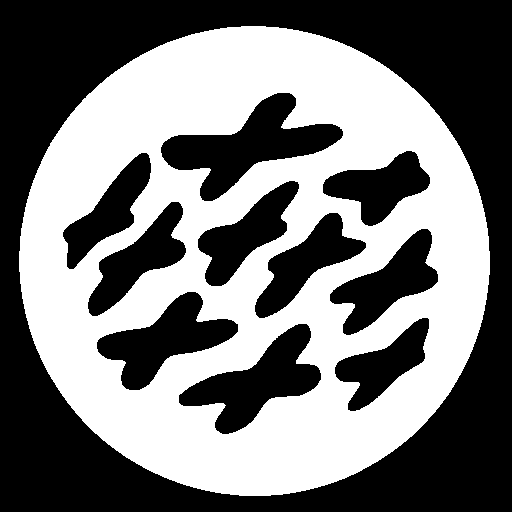}
        \captionsetup{justification=centering}
        \caption{Full angle FBP reconstruction.}
        \label{fig:fullanglefbp}
    \end{subfigure}
    \hspace{5mm}
    \begin{subfigure}[t]{.25\textwidth}
        \centering
        \includegraphics[height=2.5cm]{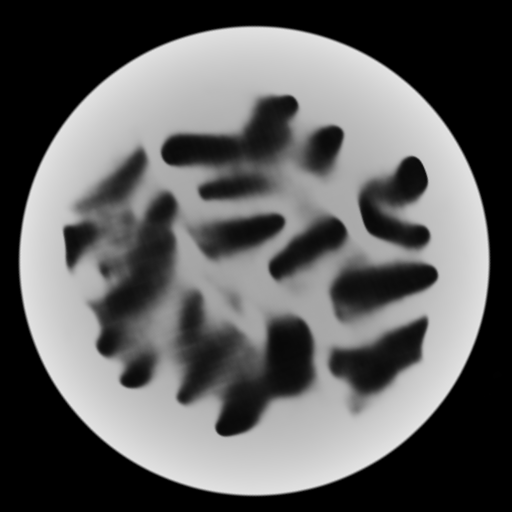}
        \captionsetup{justification=centering}
        \caption{Our limited angle reconstruction.}
        \label{fig:ourrecon}
    \end{subfigure}
    \hspace{5mm}
    \begin{subfigure}[t]{.25\textwidth}
        \centering
        \includegraphics[height=2.5cm]{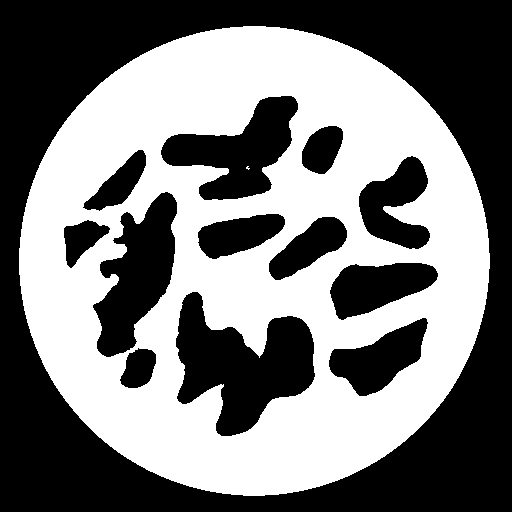}
        \captionsetup{justification=centering}
        \caption{Thresholded reconstruction.}
        \label{fig:ourreconbinary}
    \end{subfigure}

    \caption{Example of a difficult shape configuration.}
    \label{fig:limitations}
\end{figure}

The difficulty of reconstruction can vary considerably depending on the angle at which the X-rays intersect the imaged surface.
Areas where rays only travel perpendicular to the surfaces pose a particular challenge.
These areas are often referred to as \textit{invisible singularities}~\cite{Frikel_2013, doi:10.1137/0524069}.
Their prevalence increases as the angular range is reduced.
The configuration of shapes in the images of the last stage of the challenge (Figure~\ref{fig:fullanglefbp}) have been chosen in such a way as to inhibit reconstruction (Figure~\ref{fig:ourreconbinary}). In addition, the shapes are highly concave, while our synthetic dataset was constructed from mostly convex and a few moderately concave shapes. \C{For practical applications, it would be advisable to include shapes in the training dataset that are as similar as possible to the shapes in the test dataset to enhance reconstruction quality.}

\section{Summary \& Conclusion}

This paper proposes a limited-angle tomography reconstruction method based on deep end-to-end learning on synthetic data.
Even though the employed neural network was trained entirely on synthetic data, it achieves the best performance on the real world test dataset of the Helsinki Tomography Challenge of 2022.
In addition to outstanding reconstruction quality, the proposed method also shows excellent performance, taking only a few milliseconds to reconstruct an image from a limited angle sinogram on a consumer graphics card.
Furthermore, an extensive ablation study is presented, analyzing interesting aspects of the generation of the synthetic dataset and showing how the quality of the reconstruction can be improved even further.
We conclude that the proposed method has the potential to be used in various applications where the images to be reconstructed can be generated synthetically and where run-time performance is critical.

\bibliographystyle{plain}
\bibliography{bibliography}

\begin{thebibliography}{10}

\bibitem{adler2018learned}
Jonas Adler and Ozan Öktem.
\newblock Learned primal-dual reconstruction.
\newblock {\em IEEE Transactions on Medical Imaging}, 37(6):1322--1332, 2018.

\bibitem{barutcu2021limited}
Semih Barutcu, Selin Aslan, Aggelos~K. Katsaggelos, and Do{\u{g}}a Gürsoy.
\newblock Limited-angle computed tomography with deep image and physics priors.
\newblock {\em Scientific Reports}, 11(1), sep 2021.

\bibitem{brooks1976beam}
R~A Brooks and G~Di Chiro.
\newblock Beam hardening in {X}-ray reconstructive tomography.
\newblock {\em Physics in Medicine and Biology}, 21(3):390, 1976.

\bibitem{bubba2019learning}
Tatiana~A Bubba, Gitta Kutyniok, Matti Lassas, Maximilian März, Wojciech
  Samek, Samuli Siltanen, and Vignesh Srinivasan.
\newblock Learning the invisible: a hybrid deep learning-shearlet framework for
  limited angle computed tomography.
\newblock {\em Inverse Problems}, 35(6):064002, may 2019.

\bibitem{dong2019deep}
Jianbing Dong, Jian Fu, and Zhao He.
\newblock A deep learning reconstruction framework for {X}-ray computed
  tomography with incomplete data.
\newblock {\em PLoS ONE}, 14(11):e0224426, 2019.

\bibitem{douarre2018transfer}
Clément Douarre, Richard Schielein, Carole Frindel, Stefan Gerth, and David
  Rousseau.
\newblock Transfer learning from synthetic data applied to soil–root
  segmentation in x-ray tomography images.
\newblock {\em Journal of Imaging}, 4(5):65, May 2018.

\bibitem{texturing}
David~S. Ebert, F.~Kenton Musgrave, Darwyn Peachey, Ken Perlin, and Steven
  Worley.
\newblock {\em Texturing and Modeling: A Procedural Approach}.
\newblock Morgan Kaufmann Publishers Inc., San Francisco, CA, USA, 3rd edition,
  2002.

\bibitem{feldkamp1984practical}
L.~A. Feldkamp, L.~C. Davis, and J.~W. Kress.
\newblock Practical cone-beam algorithm.
\newblock {\em Journal of the Optical Society of America}, 1(6):612--619, 1984.

\bibitem{Frikel_2013}
Jürgen Frikel and Eric~Todd Quinto.
\newblock Characterization and reduction of artifacts in limited angle
  tomography.
\newblock {\em Inverse Problems}, 29(12):125007, 2013.

\bibitem{genzel2022near}
Martin Genzel, Ingo G{\"u}hring, Jan Macdonald, and Maximilian M{\"a}rz.
\newblock Near-exact recovery for tomographic inverse problems via deep
  learning.
\newblock In Kamalika Chaudhuri, Stefanie Jegelka, Le~Song, Csaba Szepesvari,
  Gang Niu, and Sivan Sabato, editors, {\em Proceedings of the 39th
  International Conference on Machine Learning}, volume 162 of {\em Proceedings
  of Machine Learning Research}, pages 7368--7381. PMLR, 17--23 Jul 2022.

\bibitem{goodfellow2020generative}
Ian Goodfellow, Jean Pouget-Abadie, Mehdi Mirza, Bing Xu, David Warde-Farley,
  Sherjil Ozair, Aaron Courville, and Yoshua Bengio.
\newblock Generative adversarial networks.
\newblock {\em Commun. ACM}, 63(11):139–144, 2020.

\bibitem{goy2019high}
Alexandre Goy, Girish Rughoobur, Shuai Li, Kwabena Arthur, Akintunde~I.
  Akinwande, and George Barbastathis.
\newblock High-resolution limited-angle phase tomography of dense layered
  objects using deep neural networks.
\newblock {\em Proceedings of the National Academy of Sciences},
  116(40):19848--19856, September 2019.

\bibitem{he2016deep}
Kaiming He, Xiangyu Zhang, Shaoqing Ren, and Jian Sun.
\newblock Deep residual learning for image recognition.
\newblock In {\em 2016 IEEE Conference on Computer Vision and Pattern
  Recognition (CVPR)}, pages 770--778, 2016.

\bibitem{hendrycks2016gaussian}
Dan Hendrycks and Kevin Gimpel.
\newblock Gaussian error linear units ({GELUs}).
\newblock {\em arXiv preprint arXiv:1606.08415}, 2016.

\bibitem{hounsfield1973computerized}
Godfrey~N Hounsfield.
\newblock Computerized transverse axial scanning (tomography): Part 1.
  description of system.
\newblock {\em The British journal of radiology}, 46(552):1016--1022, 1973.

\bibitem{inouye1979image}
Tamon Inouye.
\newblock Image reconstruction with limited angle projection data.
\newblock {\em IEEE Transactions on Nuclear Science}, 26(2):2665--2669, 1979.

\bibitem{ioffe2015batch}
Sergey Ioffe and Christian Szegedy.
\newblock Batch normalization: Accelerating deep network training by reducing
  internal covariate shift.
\newblock In {\em International Conference on Machine Learning}, pages
  448--456. PMLR, 2015.

\bibitem{jin2017deep}
Kyong~Hwan Jin, Michael~T. McCann, Emmanuel Froustey, and Michael Unser.
\newblock Deep convolutional neural network for inverse problems in imaging.
\newblock {\em IEEE Transactions on Image Processing}, 26(9):4509--4522, 2017.

\bibitem{kingma2014adam}
Diederik~P Kingma and Jimmy Ba.
\newblock Adam: A method for stochastic optimization.
\newblock {\em arXiv preprint arXiv:1412.6980}, 2014.

\bibitem{lee2018deep}
Hoyeon Lee, Jongha Lee, Hyeongseok Kim, Byungchul Cho, and Seungryong Cho.
\newblock Deep-neural-network-based sinogram synthesis for sparse-view {CT}
  image reconstruction.
\newblock {\em IEEE Transactions on Radiation and Plasma Medical Sciences},
  3(2):109--119, 2019.

\bibitem{opengl}
Bill~M Licea-Kane, Randi~J Rost, Dan Ginsburg, John~M Kessenich, Barthold
  Lichtenbelt, Hugh Malan, and Mike Weiblen.
\newblock {\em {OpenGL} Shading Language}.
\newblock Addison-Wesley Educational, Boston, MA, 3 edition, July 2009.

\bibitem{liu2022convnet}
Zhuang Liu, Hanzi Mao, Chao-Yuan Wu, Christoph Feichtenhofer, Trevor Darrell,
  and Saining Xie.
\newblock A convnet for the 2020s.
\newblock In {\em 2022 IEEE/CVF Conference on Computer Vision and Pattern
  Recognition (CVPR)}, pages 11966--11976, 2022.

\bibitem{mcleavy2021future}
CM~McLeavy, MH~Chunara, RJ~Gravell, A~Rauf, A~Cushnie, C~Staley Talbot, and
  RM~Hawkins.
\newblock The future of {CT}: deep learning reconstruction.
\newblock {\em Clinical radiology}, 76(6):407--415, 2021.

\bibitem{meaney2022helsinki}
Alexander Meaney, Fernando Silva~de Moura, and Samuli Siltanen.
\newblock {Helsinki Tomography Challenge 2022 open tomographic dataset ({HTC}
  2022)}, 2022.

\bibitem{doi:10.1137/1.9781611972344}
Jennifer~L. Mueller and Samuli Siltanen.
\newblock {\em Linear and Nonlinear Inverse Problems with Practical
  Applications}.
\newblock Society for Industrial and Applied Mathematics, Philadelphia, PA,
  2012.

\bibitem{niklason1997digital}
L~T Niklason, B~T Christian, L~E Niklason, D~B Kopans, D~E Castleberry, B~H
  Opsahl-Ong, C~E Landberg, P~J Slanetz, A~A Giardino, R~Moore, D~Albagli, M~C
  DeJule, P~F Fitzgerald, D~F Fobare, B~W Giambattista, R~F Kwasnick, J~Liu,
  S~J Lubowski, G~E Possin, J~F Richotte, C~Y Wei, and R~F Wirth.
\newblock Digital tomosynthesis in breast imaging.
\newblock {\em Radiology}, 205(2):399--406, 1997.
\newblock PMID: 9356620.

\bibitem{paschalis2004tomographic}
P~Paschalis, N.D Giokaris, A~Karabarbounis, G.K Loudos, D~Maintas, C.N
  Papanicolas, V~Spanoudaki, Ch~Tsoumpas, and E~Stiliaris.
\newblock Tomographic image reconstruction using artificial neural networks.
\newblock {\em Nuclear Instruments and Methods in Physics Research Section A:
  Accelerators, Spectrometers, Detectors and Associated Equipment},
  527(1):211--215, 2004.
\newblock Proceedings of the 2nd International Conference on Imaging
  Technologies in Biomedical Sciences.

\bibitem{doi:10.1137/0524069}
Eric~Todd Quinto.
\newblock Singularities of the {X}-ray transform and limited data tomography in
  $\mathbb{R}^2$ and $\mathbb{R}^3$.
\newblock {\em SIAM Journal on Mathematical Analysis}, 24(5):1215--1225, 1993.

\bibitem{riba2020kornia}
Edgar Riba, Dmytro Mishkin, Daniel Ponsa, Ethan Rublee, and Gary Bradski.
\newblock Kornia: an open source differentiable computer vision library for
  {P}y{T}orch.
\newblock In {\em 2020 IEEE Winter Conference on Applications of Computer
  Vision (WACV)}, pages 3663--3672, 2020.

\bibitem{ronneberger2015u}
Olaf Ronneberger, Philipp Fischer, and Thomas Brox.
\newblock U-net: Convolutional networks for biomedical image segmentation.
\newblock In Nassir Navab, Joachim Hornegger, William~M. Wells, and
  Alejandro~F. Frangi, editors, {\em Medical Image Computing and
  Computer-Assisted Intervention -- MICCAI 2015}, pages 234--241, Cham, 2015.
  Springer International Publishing.

\bibitem{schwab2019deep}
Johannes Schwab, Stephan Antholzer, and Markus Haltmeier.
\newblock Deep null space learning for inverse problems: convergence analysis
  and rates.
\newblock {\em Inverse Problems}, 35(2):025008, jan 2019.

\bibitem{shepp1974fourier}
L.~A. Shepp and B.~F. Logan.
\newblock The fourier reconstruction of a head section.
\newblock {\em IEEE Transactions on Nuclear Science}, 21(3):21--43, 1974.

\bibitem{sidky2022report}
Emil~Y. Sidky and Xiaochuan Pan.
\newblock Report on the {AAPM} deep-learning sparse-view {CT} grand challenge.
\newblock {\em Medical Physics}, 49(8):4935--4943, February 2022.

\bibitem{van2016fast}
Wim van Aarle, Willem~Jan Palenstijn, Jeroen Cant, Eline Janssens, Folkert
  Bleichrodt, Andrei Dabravolski, Jan~De Beenhouwer, K.~Joost Batenburg, and
  Jan Sijbers.
\newblock Fast and flexible {X}-ray tomography using the {ASTRA} toolbox.
\newblock {\em Optics Express}, 24(22):25129--25147, 2016.

\bibitem{van2015astra}
Wim Van~Aarle, Willem~Jan Palenstijn, Jan De~Beenhouwer, Thomas Altantzis, Sara
  Bals, K~Joost Batenburg, and Jan Sijbers.
\newblock The {ASTRA} toolbox: A platform for advanced algorithm development in
  electron tomography.
\newblock {\em Ultramicroscopy}, 157:35--47, 2015.

\bibitem{wurfl2016deep}
Tobias W{\"u}rfl, Florin~C. Ghesu, Vincent Christlein, and Andreas Maier.
\newblock Deep learning computed tomography.
\newblock In Sebastien Ourselin, Leo Joskowicz, Mert~R. Sabuncu, Gozde Unal,
  and William Wells, editors, {\em Medical Image Computing and
  Computer-Assisted Intervention - MICCAI 2016}, pages 432--440, Cham, 2016.
  Springer International Publishing.

\bibitem{wurfl2018deep}
Tobias Würfl, Mathis Hoffmann, Vincent Christlein, Katharina Breininger, Yixin
  Huang, Mathias Unberath, and Andreas~K. Maier.
\newblock Deep learning computed tomography: Learning projection-domain weights
  from image domain in limited angle problems.
\newblock {\em IEEE Transactions on Medical Imaging}, 37(6):1454--1463, 2018.

\bibitem{yang2020tomographic}
Xiaogang Yang, Maik Kahnt, Dennis Br{\"u}ckner, Andreas Schropp, Yakub Fam,
  Johannes Becher, J-D Grunwaldt, Thomas~L Sheppard, and Christian~G Schroer.
\newblock Tomographic reconstruction with a generative adversarial network.
\newblock {\em Journal of Synchrotron Radiation}, 27(2):486--493, 2020.

\bibitem{yim2021limited}
Dobin Yim, Burnyoung Kim, and Seungwan Lee.
\newblock Limited-angle {CT} reconstruction via data-driven deep neural
  network.
\newblock In Hilde Bosmans, Wei Zhao, and Lifeng Yu, editors, {\em Medical
  Imaging 2021: Physics of Medical Imaging}. {SPIE}, February 2021.

\bibitem{zhang2020artifact}
Qiyang Zhang, Zhanli Hu, Changhui Jiang, Hairong Zheng, Yongshuai Ge, and Dong
  Liang.
\newblock Artifact removal using a hybrid-domain convolutional neural network
  for limited-angle computed tomography imaging.
\newblock {\em Physics in Medicine \& Biology}, 65(15):155010, jul 2020.

\bibitem{zhu2018image}
Bo~Zhu, Jeremiah~Z. Liu, Stephen~F. Cauley, Bruce~R. Rosen, and Matthew~S.
  Rosen.
\newblock Image reconstruction by domain-transform manifold learning.
\newblock {\em Nature}, 555(7697):487--492, March 2018.

\end{thebibliography}

\end{document}